%% file: draft_prl.tex
\pdfoutput=1

\documentclass[prl,superscriptaddress,amsmath,amssymb,twocolumn]{revtex4-2}

\usepackage{graphicx,bm,amsmath}
\usepackage[usenames,dvipsnames]{xcolor}
\usepackage[colorlinks,bookmarks=false,citecolor=NavyBlue,linkcolor=Red,urlcolor=blue,
]{hyperref}
\usepackage{physics}
\usepackage[percent]{overpic}
\usepackage[bbgreekl]{mathbbol}
\usepackage{xcolor}
\usepackage[normalem]{ulem}
\usepackage{algorithm}
\usepackage{braket}
\usepackage{algpseudocode}
\usepackage{comment}
\usepackage{amsthm}
\usepackage{enumerate}

\usepackage{algorithm}
\usepackage{algpseudocode}

\def\doi{http://dx.doi.org/}

\newcommand{\be}{\begin{equation}}
\newcommand{\ee}{\end{equation}}
\newcommand{\bec}{\begin{equation*}}
\newcommand{\eec}{\end{equation*}}
\newcommand{\bea}{\begin{eqnarray}}
\newcommand{\eea}{\end{eqnarray}}

\newcommand{\titleinfo}{Nonstabilizerness via matrix product states in the Pauli basis}

\begin{document}
\title{\titleinfo}

\author{Poetri Sonya Tarabunga}
\affiliation{The Abdus Salam International Centre for Theoretical Physics (ICTP), Strada Costiera 11, 34151 Trieste, Italy}
\affiliation{SISSA, Via Bonomea 265, 34136 Trieste, Italy}
\affiliation{INFN, Sezione di Trieste, Via Valerio 2, 34127 Trieste, Italy}

\author{Emanuele Tirrito}
\affiliation{The Abdus Salam International Centre for Theoretical Physics (ICTP), Strada Costiera 11, 34151 Trieste, Italy}
\affiliation{Pitaevskii BEC Center, CNR-INO and Dipartimento di Fisica, Università di Trento, Via Sommarive 14, Trento, I-38123, Italy}
\author{Mari Carmen Ba\~nuls}
\affiliation{Max-Planck-Institut f\"ur Quantenoptik, Hans-Kopfermann-Straße 1, D-85748 Garching, Germany} 
\affiliation{Munich Center for Quantum Science and Technology (MCQST), Schellingstraße 4, D-80799 Munich, Germany}
\author{Marcello Dalmonte}
\affiliation{The Abdus Salam International Centre for Theoretical Physics (ICTP), Strada Costiera 11, 34151 Trieste, Italy}
\affiliation{SISSA, Via Bonomea 265, 34136 Trieste, Italy}

\begin{abstract}
Nonstabilizerness, also known as ``magic'', stands as a crucial resource for achieving a potential advantage in quantum computing. Its connection to many-body physical phenomena is poorly understood at present, mostly due to a lack of practical methods to compute it at large scales. We present a novel approach for the evaluation of nonstabilizerness within the framework of matrix product states (MPS), based on expressing the MPS directly in the Pauli basis. Our framework provides a powerful tool for efficiently calculating various measures of nonstabilizerness, including stabilizer Rényi entropies, stabilizer nullity, and Bell magic, and enables the learning of the stabilizer group of an MPS. 
We showcase the efficacy and versatility of our method in the ground states of Ising and XXZ spin chains, as well as in circuits dynamics that has recently been realized in Rydberg atom arrays, where we provide concrete benchmarks for future experiments on logical qubits up to twice the sizes already realized. 

\end{abstract}

\maketitle

\paragraph{Introduction.---}
 Quantum computers hold promise for simulating quantum systems \cite{preskill2012quantum}, a task that classical computers struggle with \cite{shor1999polynomial}. While entanglement is necessary to achieve this goal \cite{amico2008entanglement,eisert2010colloquium,cirac2012goals,bloch2012quantum,aspuru2012photonic,houck2012chip,vandersypen2005nmr,plenio2014introduction}, it is not sufficient. For example, consider the Clifford circuits, that are generated by the Hadamard $(H)$ gate, the phase $(S)$ gate, and the controlled-not $(CNOT)$ gate. The class of states that can be generated by Clifford circuits are called stabilizer states, which can be highly entangled, and yet they can be efficiently simulated classically \cite{nielsen2002quantum, gottesman1997stabilizer, PhysRevA.57.127,aaronson2004improved,PhysRevA.54.1862,PhysRevA.57.127}. In this context, the so-called nonstabilizer states or magic states \cite{bravyi2005UniversalQuantumComputation,bravyi2012magic,campbell2017roads,harrow2017quantum,veitch2014resource,wootters1987wigner,gross2006hudson,veitch2012negative,wang2019quantifying} are crucial to unlock potential quantum advantage. They enable the distillation of non-Clifford gates, such as $T$\,$=$\,$\rm{diag}(1,e^{i\pi/4})$ or $CCZ$ (controlled-controlled-Z) gate , which, combined with the Clifford gates, creates a universal gate set.

The degree to which a quantum state deviates from stabilizer states is called \textit{nonstabilizerness} or magic \cite{bravyi2005UniversalQuantumComputation}. An important task is to quantify the amount of nonstabilizerness of a given state. For example, in quantum circuits, applying non-Clifford gates increases the nonstabilizerness. Quantifying this growth is essential for understanding a circuit's potential for quantum advantage, as recent experiments demonstrate \cite{Bluvstein2023}. Moreover, for ground states of Hamiltonians, nonstabilizerness is related to the amount of non-Clifford gates needed to prepare the state. 

Much like entanglement, nonstabilizerness has been quantified within the framework of resource theory using measures of nonstabilizerness \cite{chitambar2019quantum}, and several measures have been proposed \cite{wigner1997quantum,gross2007non,veitch2012negative,wootters1987wigner,hudson1974wigner,buvzek1992superpositions,veitch2014resource,bravyi2016improved,bravyi2016trading,howard2017robustness,heinrich2019robustness,wang2020efficiently,heimendahl2021stabilizer}.  
However, most of these quantifiers are difficult to evaluate even numerically. This has hindered the study of nonstabilizerness beyond a few qubits, posing a major challenge in the field. Recently, two computable measures have been introduced: Bell magic~\cite{PRXQuantum.4.010301} and stabilizer Rényi entropies (SREs)~\cite{leone2022stabilizer}. Unfortunately, their calculation still requires evaluating an exponential number of terms. Nevertheless, notable progress has been made in their experimental measurements~\cite{PRXQuantum.4.010301,Oliviero2022, niroula2023phase, haug2023efficient, Bluvstein2023}. Moreover, several methods have been put forward to compute the SREs numerically, based on, e.g., tensor networks \cite{haug2023stabilizer,haug2023quantifying,Lami2023,tarabunga2023manybody} , Monte Carlo sampling of wavefunctions \cite{tarabunga2023magic}, and average over Clifford orbits \cite{tirrito2023,turkeshi2023measuring}. These methods have enabled the study of nonstabilizerness in many-body contexts \cite{liu2022many}, particularly its connection to criticality \cite{haug2023quantifying, tarabunga2023manybody,tarabunga2023critical,tarabunga2023magic,white2021}. However, these approaches still face limitations in their applicability and computational efficiency, especially in terms of demonstrated accessible quantities.

\begin{figure*}
    \centering
    \includegraphics[width=1.0\linewidth]{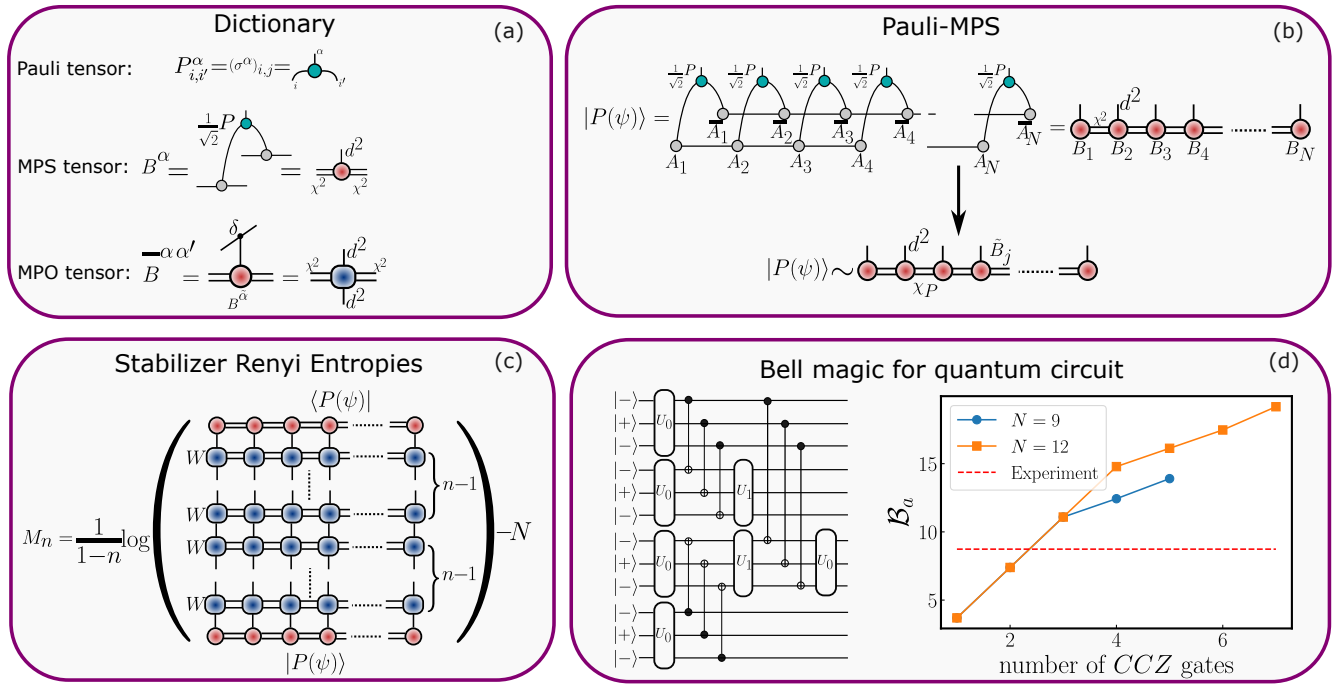}
    \vspace*{-6mm}
    \caption{
    (a) Definitions of tensors used for the construction of Pauli-MPS. (b) Construction of Pauli-MPS. (c) The SRE represented as the contraction of a two-dimensional tensor network. (d) The additive Bell magic in a scrambling circuit recently experimentally realized in Ref. \cite{Bluvstein2023}. The red dashed line indicates the highest value of the additive Bell magic experimentally measured in Ref. \cite{Bluvstein2023}.
    }
    \label{fig:folding_mps}
\end{figure*}

In this work, we demonstrate how, for states represented by matrix product states (MPS), several nonstabilizerness measures can be cast in the language of tensor networks (TN)~\cite{Verstraete2008,Schollwoeck2011,Orus2014annphys,Silvi2019tn,Okunishi2022,Banuls2023}, whose contractions can be approximated using standard algorithms. More concretely, we represent the Pauli spectrum of the state as an MPS, cf. Fig. \ref{fig:folding_mps} (a,b). 
This allows one to compute not only the SRE, but also Bell magic, which has so far not been quantified in large systems, as it is too costly to compute by existing methods. For the SRE in particular, we express it as a two-dimensional tensor network as shown in Fig. \ref{fig:folding_mps} (c), thereby enabling approximate contraction using established MPS methods. Furthermore, we explain how our framework leads to efficient computation of stabilizer nullity, a genuine nonstabilizerness monotone, which in turn allows us to identify the stabilizer group of the state. We benchmark our method through various examples, including the quantum Ising chain, the XXZ chain, and random Clifford circuits with nonstabilizer states input. We further applied our method to compute Bell magic in a scrambling circuit (see Fig. \ref{fig:folding_mps} (d)) recently experimentally implemented in Rydberg atom arrays \cite{Bluvstein2023}. Reaching system sizes beyond the current experimental capabilities, our method can thus be used to verify and benchmark future experiments.

\paragraph{MPS in the Pauli basis.---}
Let us consider a system of $N$ qubits in a pure state $|\psi \rangle$ given by an MPS of bond dimension $\chi$:
\be \label{eq:mps}
|\psi \rangle=\sum_{s_1,s_2,\cdots,s_N} A^{s_1}_1 A^{s_2}_2 \cdots A^{s_N}_N |s_1,s_2,\cdots s_N \rangle
\ee
with $A_i^{s_i}$ being $\chi$\,$\times$\,$\chi$ matrices, except at the left (right)
boundary where $A_1^{s_1}$ ($A_N^{s_N}$) is a $1$\,$\times$\,$\chi$ ($\chi$\,$\times$\,$1$) row (column) vector. Here $s_i$\,$\in$\,$\left \lbrace 0, 1 \right \rbrace$  is a local computational basis. The state is assumed right-normalised, namely $\sum_{s_i} A_i^{s_i \dagger} A_i^{s_i}$\,$=$\,$1$. 
Let us define the binary string $\boldsymbol{\alpha}$\,$=$\,$(\alpha_1, \cdots, \alpha_N)$ with $\alpha_j$\,$\in$\,$\{00,01,10,11\}$. The Pauli strings are defined as $P_{\boldsymbol{\alpha}} $\,$=$\,$ P_{\alpha_1}  \otimes P_{\alpha_2} \otimes \cdots  \otimes P_{\alpha_N}$
where $P_{00}$\,$=$\,$I, P_{01}$\,$=$\,$\sigma^x, P_{11}$\,$=$\,$\sigma^y,$ and $P_{10}$\,$=$\,$\sigma^z$. We define the Pauli vector of $|\psi \rangle$ as $|P(\psi) \rangle$ with elements $  \langle \boldsymbol{\alpha} |P(\psi) \rangle$\,$=$\,$ \langle \psi | P_{\boldsymbol{\alpha}} |\psi \rangle / \sqrt{2^N}$. Also known as the Pauli spectrum~\cite{Beverland2020}, this was recently studied in the context of many-body systems~\cite{turkeshi2023pauli}. When $|\psi \rangle$ has an MPS structure as in Eq. \eqref{eq:mps}, the Pauli vector can also be expressed as an MPS as follows
\be  \label{eq:pauli_mps}
|P(\psi) \rangle=\sum_{\alpha_1,\alpha_2,\cdots,\alpha_N} B^{\alpha_1}_1 B^{\alpha_2}_2 \cdots B^{\alpha_N}_N | \alpha_1, \cdots, \alpha_N \rangle
\ee
where $B^{\alpha_i}_i $\,$=$\,$ \sum_{s,s'} \langle s | P_{\alpha_i} | s' \rangle A^s_i \otimes \overline{A^{s'}_i} / \sqrt{2}$ are $\chi^2 \times \chi^2$ matrices, as shown in Fig. \ref{fig:folding_mps}.
Note that the MPS is normalized due to the relation $\frac{1}{2^N} \sum_{\boldsymbol{\alpha}} \langle \psi | P_{\boldsymbol{\alpha}} | \psi \rangle^2 $\,$=$\,$ 1$
which holds for pure states. Moreover, it retains the right normalization, due to the identity $\frac{1}{2} \sum_\alpha P_\alpha (\cdot) P_\alpha $\,$=$\,$ \mathbb{1} \Tr[\cdot] $. Consequently, the entanglement spectrum of $|P(\psi) \rangle$ is given by $\lambda'_{i,j}$\,$=$\,$ \lambda_i \lambda_j$ for $i,j$\,$=$\,$1,2,\cdots,\chi$, where $\lambda_i$ is the entanglement spectrum of $|\psi \rangle$, and hence the von Neumann entropy is doubled. Note also that the coefficients of $|P(\psi) \rangle$ in the Pauli basis \eqref{eq:pauli_mps} are real, since the Pauli operators are Hermitian for spin-1/2 systems, although the local tensors $B_i$ are not necessarily real.

Since the Pauli operators provide an orthonormal basis in the space of Hermitian operators, one can expand the density matrix  as $|\psi \rangle \langle \psi |$\,$=$\,$\frac{1}{2^N} \sum_\mathbf{\alpha} \langle \psi | P_{\boldsymbol{\alpha}} |\psi \rangle P_{\boldsymbol{\alpha}}$. Therefore, the Pauli spectrum is simply the coefficients of $|\psi \rangle \langle \psi |$ in the basis of Pauli operators, i.e., the Pauli basis. 
As we show below, the MPS representation in the Pauli basis provides a powerful and versatile tool to compute various measures of nonstabilizerness. 
 Specifically, we will consider the measures SRE \cite{leone2022stabilizer}, stabilizer nullity \cite{Beverland2020}, and Bell magic \cite{PRXQuantum.4.010301}.   

The SRE is defined as \cite{leone2022stabilizer}
\be \label{eq:SRE}
M_n \left( | \psi \rangle \right)= \frac{1}{1-n} \log_2 \left \lbrace \sum_{\boldsymbol{\alpha}} \frac{|\langle \psi | P_{\boldsymbol{\alpha}} | \psi \rangle|^{2n}}{2^N} \right \rbrace \,  . 
\ee
 Then, the additive Bell magic $\mathcal{B}_a \left( | \psi \rangle \right)$ is given by
\be \label{eq:bell_magic}
\mathcal{B}_a \left( | \psi \rangle \right) = - \log_2 \left( 1- \mathcal{B} \left( | \psi \rangle \right) \right) \, ,
\ee
where $\mathcal{B}\left( | \psi \rangle \right) $ is the Bell magic, defined as \cite{PRXQuantum.4.010301}
\be
\mathcal{B} \left( | \psi \rangle \right) = {\sum_{\scriptsize
    \begin{array}{l} \boldsymbol{\alpha},\boldsymbol{\alpha}^{\prime}, \\ \boldsymbol{\beta},\boldsymbol{\beta}^{\prime} 
    \end{array}
    }}
    \Xi(\boldsymbol{\alpha}) \Xi(\boldsymbol{\alpha}^{\prime}) \Xi(\boldsymbol{\beta}) \Xi(\boldsymbol{\beta}^{\prime}) || [P_{\boldsymbol{\alpha} \oplus \boldsymbol{\alpha}^{\prime}},P_{\boldsymbol{\beta} \oplus \boldsymbol{\beta}^{\prime}}] ||_{\infty},
\ee
where $\Xi(\boldsymbol{\alpha})$\,$=$\,$|\langle \psi | P_{\boldsymbol{\alpha}} | \psi \rangle|^{2}/2^N$ is known as the characteristic function \cite{Gross2021}, and $\oplus$ denotes a bit-wise XOR \cite{footnote1}. The infinity norm is zero when the Pauli strings commute and 2 otherwise. Finally, the stabilizer nullity $\nu(| \psi \rangle)$ is simply related to the size of the stabilizer group ${\rm Stab}(\psi)$, which is the group of Pauli strings that stabilize $| \psi \rangle$. The stabilizer nullity is defined as \cite{Beverland2020} 
\begin{equation}
    \nu(| \psi \rangle) = N - \log_2 \left( |{\rm Stab}(\psi)| \right).
\end{equation}
More details on these measures can be found in the Supplemental Material \cite{supmat}. Hereafter, we will drop the dependence on $|\psi \rangle$ to keep the notation light.

\paragraph{Replica Pauli-MPS.---}
The replica method in MPS was introduced to compute the SRE of MPS in Ref.~\cite{haug2023quantifying}. While exact, for practical purposes, the original formulation performed inferiorly with respect to Pauli sampling methods due to the extremely high cost with respect to the bond dimension \cite{Lami2023,haug2023stabilizer,tarabunga2023manybody}. Indeed, evaluating the SRE for an integer index $n>1$ required a computational cost of $O\left(\chi^{6n}\right)$, rendering it impractical for even the simplest case $n$\,$=$\,$2$, where previous computations were restricted to $\chi$\,$=$\,$12$~\cite{haug2023quantifying,haug2023stabilizer}. Note that sampling methods have their own limitations, as the number of samples has to scale exponentially~\cite{footnote8}. Here, we show that the MPS in the Pauli basis can be exploited to significantly reduce the cost of the replica trick, opening doors for its use in practical application and making it superior also compared to sampling methods in terms of computational efficiency and scalability.

To do so, we define a diagonal operator $W$ whose diagonal elements are the components of the Pauli vector, $\langle  \boldsymbol{\alpha'}| W |\boldsymbol{\alpha} \rangle= \delta_{\boldsymbol{\alpha'},\boldsymbol{\alpha}} \langle \boldsymbol{\alpha'} |P(\psi) \rangle$. The MPO form of $W$ reads
\begin{equation}
    W =\sum_{\boldsymbol{\alpha},\boldsymbol{\alpha'}} \overline{B}^{\alpha_1,\alpha'_1}_1 \overline{B}^{\alpha_2,\alpha'_2}_2 \cdots \overline{B}^{\alpha_N,\alpha'_N}_N | \alpha_1, \cdots, \alpha_N \rangle \langle \alpha'_1, \cdots, \alpha'_N |
\end{equation}
where $\overline{B}^{\alpha_i,\alpha'_i}_i$\,$=$\,$ B^{\alpha_i}_i \delta_{\alpha_i,\alpha'_i}$.
Applying $W$ $n$\,$-$\,$1$ times to $|P(\psi) \rangle$, we obtain $|P^{(n)}(\psi) \rangle $\,$=$\,$W^{n-1} |P(\psi) \rangle$, which is a vector with elements $\langle \boldsymbol{\alpha} |P^{(n)}(\psi) \rangle$\,$=$\,$\langle \psi | P_{\boldsymbol{\alpha}} |\psi \rangle^n / \sqrt{2^{Nn}}$. 
We denote the local tensors of $|P^{(n)}(\psi) \rangle$ by $B^{(n)\alpha_i}_i$. We have
\begin{equation}
    \frac{1}{2^{Nn}} \sum_{\boldsymbol{\alpha}} \langle \psi | P_{\boldsymbol{\alpha}} | \psi \rangle^{2n} = \langle P^{(n)}(\psi)  | P^{(n)}(\psi)  \rangle
\end{equation}
and \cite{footnote2}
\begin{equation} \label{eq:sre_replica}
    M_n = \frac{1}{1-n} \log{\langle P^{(n)}(\psi)  | P^{(n)}(\psi)  \rangle} - N.
\end{equation}

The exact bond dimension of $|P^{(n)} \rangle$ is $\min\left(\chi^{2n},4^{N/2}\right)$, i.e., for large systems it grows exponentially with the order $n$, as the cost in Ref.~\cite{haug2023quantifying}. However, by interpreting it as the repeated application of a MPO $W$ onto an MPS, we can sequentially compress the resulting MPS after every iteration, and keep the best description of the resulting state as a MPS with some upper-bounded bond dimension $\chi_n$~ (see Supplemental Material \cite{supmat}). 
This can be done with standard tensor network routines used, e.g., in the simulation of time evolution~\cite{Garc_a_Ripoll_2006,Schollwoeck2011}. These methods allow us to monitor the error of the truncation, for example, by doing convergence analysis~\cite{footnote3}.

The Pauli-MPS itself can also be approximated with a bond dimension $\chi_P$\,$<$\,$\chi^2$. The computational cost of this compression is $O\left( \chi_P^2 \chi^2 + \chi^3 \chi_P \right)$. This is particularly advantageous for states with exponentially decaying entanglement spectrum (e.g. in gapped phases), in which case $\chi_P$ can be truncated to a value much smaller than $\chi^2$. Assuming $\chi_P$\,$\approx$\,$\chi$, this results in the overall cost of $O\left(N\chi^4\right)$. By comparison, the computational cost of direct Pauli sampling is $O\left(N N_S \chi^3\right)$ \cite{Lami2023,haug2023stabilizer}, where $N_S$ is the number of samples. Consequently, our method becomes superior compared to the latter when $N_S$\,$\gtrsim$\,$\chi$. Since $N_S$ typically grows exponentially with $N$ for the estimation of $M_2$, our method offers significant efficiency gains for large $N$, for states with bounded $\chi$. Although this is at the cost of introducing an approximation, convergence can be controlled by monitoring truncation error, a standard practice in tensor networks. On the other hand, if $\chi$ scales exponentially (e.g. in volume-law phases), our method may become too expensive compared to sampling.

\begin{figure}
        \centering
        \includegraphics[width=0.47\textwidth]{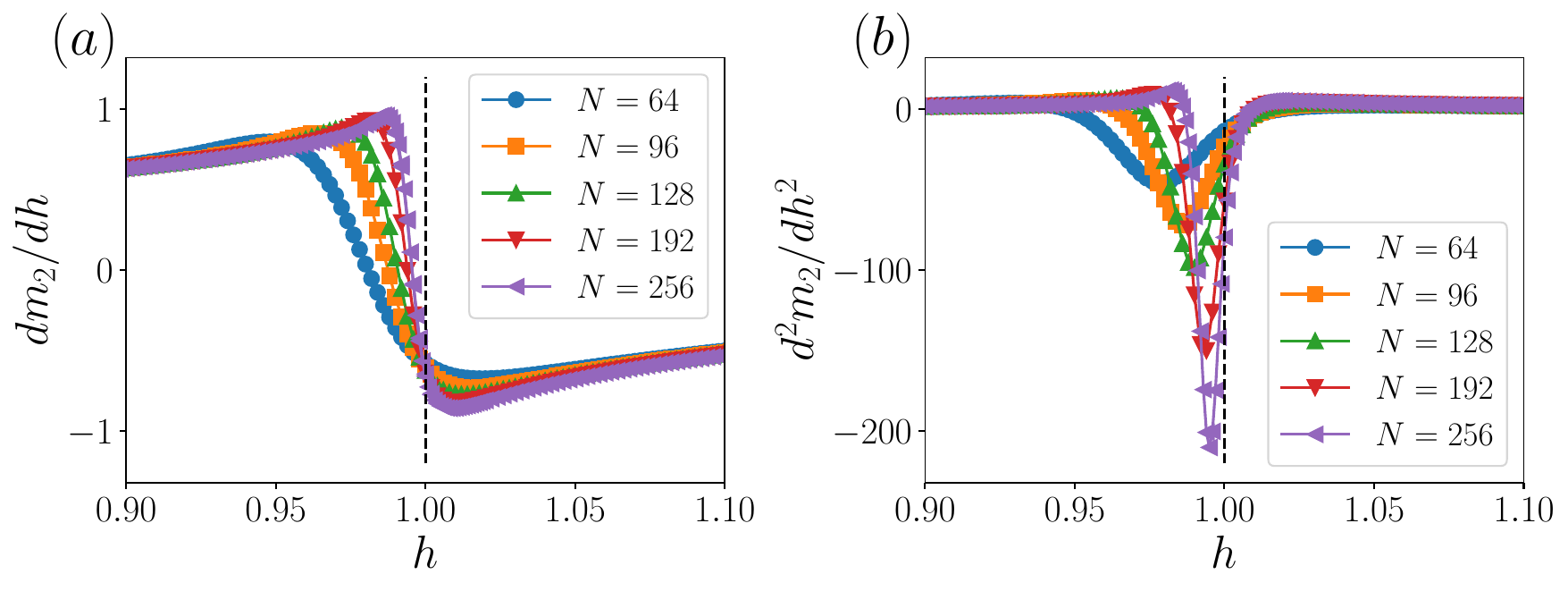}
        \vspace*{-4mm}
        \caption{(a) The first and (b) second derivative of the SRE density $m_2$\,$=$\,$M_2/N$ in the ground states of the quantum Ising chain as a function of the transverse field $h$.  }
        \label{fig:ising_derivatives}
    \end{figure}

We benchmarked the method in the XXZ chain, whose Rényi-2 SRE could not be computed accurately for $N$\,$>$\,$30$ in the previous study \cite{haug2023stabilizer}. The numerical results along with the convergence analysis are detailed in the Supplemental Material \cite{supmat}.  In addition, as a concrete application, we calculated the SRE in the quantum Ising chains, $H_{\text{Ising}}$\,$=$\,$-\sum_{\langle i,j \rangle} \sigma_i^x \sigma_{j}^x - h \sum_i \sigma^z_i$, previously considered in Refs. \cite{oliviero2022ising,haug2023quantifying, tarabunga2023manybody, Rattacaso2023}. We obtain the ground states using DMRG with $\chi$\,$=$\,$40$ and compute the SRE using replica Pauli-MPS, imposing a truncation error threshold of $\epsilon$\,$=$\,$10^{-9}$. Fig. \ref{fig:ising_derivatives} shows the derivatives of the SRE around the critical point $h$\,$=$\,$1$. We observe that the second derivative appears to diverge at the critical point, mirroring the results of Ref. \cite{tarabunga2023magic} for Rokhsar-Kivelson states. These results further solidify the role of nonstabilizerness as a useful diagnostic tool for identifying criticality in quantum systems \cite{haug2023quantifying,tarabunga2023manybody,tarabunga2023critical,tarabunga2023magic}. Note that calculating derivatives with sampling-based approaches become increasingly challenging with the derivative order due to the presence of statistical errors.

We further notice that the norm of $|P^{(n)}(\psi) \rangle$ can be interpreted as the contraction of a two-dimensional tensor network (see Fig.~\ref{fig:folding_mps} (c)). This allows for alternative strategies to perform the contraction as for example transverse contractions~\cite{PhysRevLett.102.240603,muller2012tensor,hastings2015connecting}, corner transfer matrix \cite{PhysRevB.85.205117} or tensor renormalization group (TRG) techniques \cite{PhysRevLett.99.120601}.  We leave these possibilities for future investigations (see \cite{supmat}).

\paragraph{Bell magic.---}
Next, we consider Bell magic~\cite{haug2023quantifying}, that has recently been experimentally measured in Ref.~\cite{Bluvstein2023}. 
To compute Bell magic, we first evaluate the self-convolution of $| P^{(2)}(\psi)  \rangle$:
\begin{equation}
    |Q(\psi) \rangle=\sum_{\alpha_1,\alpha_2,\cdots,\alpha_N} C^{\alpha_1}_1 C^{\alpha_2}_2 \cdots C^{\alpha_N}_N | \alpha_1, \cdots, \alpha_N \rangle
\end{equation}
where $C^{\alpha_i}_i$\,$=$\,$\sum_{\beta,\gamma} \delta_{\beta \oplus \gamma, \alpha_i} B^{(2)\beta}_i \otimes B^{(2)\gamma}_i$ \cite{footnote9}. As before, we can compress $|Q(\psi) \rangle$ to keep the computational cost manageable. Its tensor network representation can be found in the Supplemental Material \cite{supmat}. Then, the additive Bell magic is given by
\begin{equation} \label{eq:additive_bell_magic}
    \mathcal{B}_a = -\log{ \langle Q(\psi) | \Lambda \otimes \Lambda \otimes \cdots \otimes \Lambda  |Q(\psi) \rangle} 
\end{equation}
where $ \langle \alpha' | \Lambda | \alpha \rangle $\,$=$\,$1 $ if $\left[P_{\alpha} ,P_{\alpha'}  \right]$\,$=$\,$0$ and $ \langle \alpha' | \Lambda | \alpha \rangle$\,$=$\,$-1 $ otherwise. 

We have benchmarked the additive Bell magic calculations in the Ising and XXZ chains, where we find similar behavior to that of the SRE in both cases (see Supplemental Material \cite{supmat}). Furthermore, we computed Bell magic in a state prepared by a quantum circuit recently realized in Ref. \cite{Bluvstein2023}, shown in Fig. \ref{fig:folding_mps} (d).  We verify that the additive Bell magic increases as a function of the number of $CCZ$ gates applied. Similar growth can also be observed in $T$-doped random Clifford circuits (see Supplemental Material \cite{supmat}).

\paragraph{Stabilizer nullity and stabilizer group.---}
Here, we show that stabilizer nullity \cite{Beverland2020} can be calculated using MPS in the Pauli basis. The key insight is that stabilizer nullity can be expressed as a particular limit of the SRE~\cite{footnote4}:
\begin{equation} \label{eq:nullity_sre}
    \nu = \lim_{n\to \infty} (n-1)M_n.
\end{equation}
This is evident from Eq. \eqref{eq:SRE}, where taking the limit $n$\,$\to$\,$\infty$ effectively eliminates all Pauli strings except those for which $\langle\psi|P_{\boldsymbol{\alpha}}|\psi\rangle$\,$=$\,$\pm 1$, i.e., those within the stabilizer group ${\rm Stab}(\psi)$.

From Eq. \eqref{eq:nullity_sre} and Eq. \eqref{eq:sre_replica}, we see that the nullity can be obtained by applying $W$ repeatedly to $|P(\psi) \rangle$. Furthermore, one can apply the trick employed in the exponential tensor renormalization group \cite{Chen2018} to reach the large $n$ limit exponentially faster. The idea is to iteratively construct a new diagonal MPO $W_k$ after each iteration, based on the current MPS $|P_k \rangle$. More details on the algorithm can be found in the Appendix.  The time complexity is $O\left(N \log(\nu) \chi_{\mathrm{max}}^4 \right)$, where $\chi_{\mathrm{max}}$ is the maximum bond dimension across iterations \cite{footnote5}. The linear scaling with $N$ surpasses existing methods utilizing Bell difference sampling \cite{grewal2023efficient,hangleiter2023bell}, that can be applied directly to an MPS through perfect Pauli sampling \cite{Lami2023,haug2023stabilizer} with cost $O\left(N^3 + N^2 \chi^3\right)$. Our algorithm thus establishes a new state-of-the-art for states efficiently represented by MPS.

Further, we can learn the stabilizer group of $| \psi  \rangle$ by an algorithm based on Bell sampling~\cite{montanaro2017learning} (see Appendix). Learning the stabilizer group of a state has previously been used as a first step to learn the full description of states prepared with few non-Clifford gates \cite{leone2023learning,grewal2023efficient,hangleiter2023bell}. We detail how to perform this task within our MPS framework in the Supplemental Material \cite{supmat}. Note that, while the MPS form itself is already an efficient classical description of a state, the description in terms of the stabilizer group could be useful, e.g., for simulating Clifford circuits. Furthermore, with the knowledge of the stabilizer group, one can construct a symmetric MPS in the Pauli basis, potentially leading to efficient MPS simulations of Clifford circuits.

\begin{figure}
        \centering
        \includegraphics[width=0.47\textwidth]{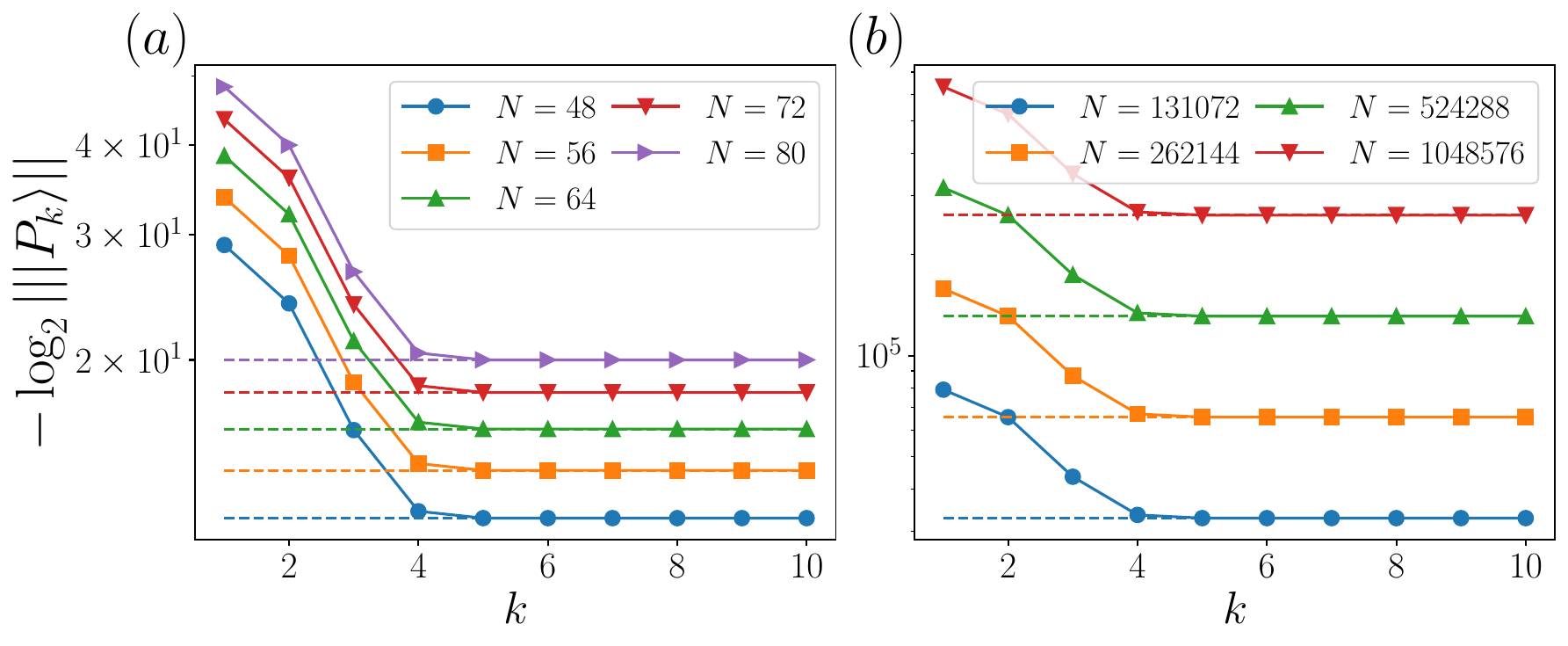}
        \vspace*{-4mm}
        \caption{We show $-\log_2 \norm{| P_k \rangle}$ at iteration $k$ in the outputs of random Clifford circuits with $N_T$\,$=$\,$N/2$ number of $T$ gates and circuit depth (a) $D$\,$=$\,$N/4$ and (b) $D$\,$=$\,$10$. After sufficiently many iterations, $-\log_2 \norm{| P_k \rangle}$ flows to $(N$\,$-$\,$\nu)/2$, denoted by the dashed lines for each system with the same color.}.
        \label{fig:nullity}
    \end{figure}

To benchmark our algorithm, we consider $T$-doped states $ |+ \rangle^{\otimes N-N_T} |T \rangle^{\otimes N_T} $, for $| + \rangle$\,$=$\,$\frac{| 0 \rangle + | 1 \rangle}{\sqrt{2}}$ and $| T \rangle$\,$=$\,$T| + \rangle$. Notably, for product states, the MPS $|P_k \rangle$ for each $k$ remains a product state, allowing for highly efficient nullity computation. We then apply a random Clifford circuit of depth $D$, which preserves the nullity $\nu$\,$=$\,$N_T$. The Clifford gates are drawn randomly from the set $\{S,H,CNOT,CZ\}$ in each layer. The two-qubit gates are applied only to nearest-neighbors. Fig. \ref{fig:nullity} (a) shows convergence of $-\log_2 \norm{| P_k \rangle}$  for $D$\,$=$\,$N/4$ and $N_T$\,$=$\,$N/2$, which according to the algorithm above should flow to $\frac{N-\nu}{2}$ as $k$\,$\to$\,$\infty$. 
For this calculation, we imposed a fixed truncation error threshold $\epsilon$\,$=$\,$10^{-6}$. 
For $N$\,$=$\,$80$, the bond dimension of $| \psi \rangle$ reaches $\chi$\,$=$\,$32$, while $\chi_P$ reaches $\chi_P$\,$=$\,$1024$. We see that convergence occurs rapidly (within 10 iterations) in all cases. For shallow circuits with constant depth $D$\,$=$\,$10$, we were able to perform simulations up to $N$\,$=$\,$1048576 \: (2^{20})$, as shown in Fig. \ref{fig:nullity} (b). Here, the maximum bond dimension of is $\chi$\,$=$\,$16$. We have verified that the computational time approximately grows linearly with $N$. The total CPU time was 1.5 days for $N$\,$=$\,$1048576$ with 10 iterations \cite{footnote6}. This demonstration represents a significant leap forward by orders of magnitude in computing genuine nonstabilizerness monotones, compared to the previous attempts \cite{heinrich2019robustness,hamaguchi2023handbook} limited to $O(10)$ qubits. These results would also be extremely challenging to reproduce using approaches based on Bell sampling \cite{grewal2023efficient,hangleiter2023bell}, due to its unfavorable scaling with system size. Additional results for spin chain systems are presented in the Supplemental Material \cite{supmat}.

\paragraph{Conclusions.---}
We have proposed a new MPS framework in the Pauli basis in order to investigate nonstabilizerness in quantum many-body systems. We discuss how several measures of nonstabilizerness, including the SREs, stabilizer nullity, and Bell magic can be efficiently approximated within our approach, and we demonstrated its usefulness in several scenarios, from ground states of spin chains to quantum circuits. Our framework can be easily generalized to mixed states and qudit systems, and it can be used to improve approaches utilizing perfect Pauli sampling \cite{supmat}. It could also be applied directly to higher-dimensional isometric tensor network states \cite{Zaletel2020}.

In terms of future investigations, it would be interesting if our MPS approach could facilitate analytical treatment of the SRE by exploiting its simple representation as a two-dimensional tensor network. Furthermore, we expect that our method would be useful to understand the role of nonstabilizerness in hybrid quantum circuits, a topic explored in recent works \cite{bejan2023dynamical,fux2023entanglementmagic}. Notably, our method allows for the efficient computation of stabilizer nullity, which is a strong monotone, and is thus suitable to characterize nonstabilizerness in such scenarios. Finally, it would be fascinating to explore the applicability of our framework to compute nonstabilizerness measures that require optimization, such as the stabilizer fidelity \cite{bravyi2019simulation} and the robustness of magic \cite{howard2017robustness}.

\begin{acknowledgments}
\paragraph{Acknowledgments.---} We thank M. Collura, M. Frau, A. Hamma,  G. Lami, M. Serbyn, and Guo-Yi Zhu, for insightful discussions. We especially thank Lorenzo Piroli for pointing out the connection between the stabilizer nullity and the SRE.
P.S.T. acknowledges support from the Simons Foundation through Award 284558FY19 to the ICTP.
M.\,D. and E.\,T. were partly supported by the MIUR Programme FARE (MEPH), by QUANTERA DYNAMITE PCI2022-132919, and by the EU-Flagship programme Pasquans2.
M.\,D. was partly supported by the PNRR MUR project PE0000023-NQSTI.
M.\,D. work was in part supported by the International Centre for Theoretical Sciences (ICTS) for participating in the program -  Periodically and quasi-periodically driven complex systems (code: ICTS/pdcs2023/6). M.C.B. was partly supported by  the DFG (German Research Foundation) under Germany's Excellence Strategy -- EXC-2111 -- 390814868; 
and by the EU-QUANTERA project TNiSQ (BA 6059/1-1).

Our numerical simulations have been performed using C++ iTensor library~\cite{itensor22}.

{\it Note added}: While completing this manuscript, we became aware of a parallel,
independent preprint on nonstabilizerness and tensor networks by Lami
and Collura [arXiv:2401.16481 (2024)], which also introduces novel
sampling methods applicable to computing stabilizer nullity.

\end{acknowledgments}

\paragraph{Appendix on stabilizer nullity and stabilizer group.---} 

The scheme to calculate the stabilizer nullity is summarized in Algorithm~\ref{alg:nullity}. We monitor the MPS norm $T_k$\,$=$\,$\norm{|P_{k} \rangle} $\,$=$\,$\sqrt{\langle P_k | P_k \rangle}$, which is related to the nullity by $\nu$\,$=$\,$\lim_{k \to \infty} N + 2 \log_2 T_k$, and convergence is achieved when the change in $T_k$ falls below a threshold $\epsilon$. The final MPS is the fixed point $| G(\psi)  \rangle$ which satisfies $W_{\infty}| G(\psi)  \rangle$\,$=$\,$\sqrt{2^{\nu-N}} | G(\psi)  \rangle $ \cite{footnote7}. The convergence is super-exponentially fast with the number of iterations, and is controlled by the ``magic gap'' \cite{Bu2023}, another nonstabilizerness measure calculable using similar approach \cite{supmat}. The time complexity is $O\left(N \log(\nu) \chi_{\mathrm{max}}^4 \right)$, where $\chi_{\mathrm{max}}$ is the maximum bond dimension across iterations \cite{footnote5}.

\begin{algorithm}[H] 
\caption{Stabilizer nullity via Pauli-MPS} \label{alg:nullity}
\begin{flushleft}
 \textbf{Input}: Pauli vector $| P(\psi) \rangle$ and threshold $\epsilon$ \\
 \textbf{Output}: Stabilizer nullity $\nu$

\end{flushleft}
\begin{algorithmic}[1]
\State $|P_0 \rangle$\,$\leftarrow$\,$|P(\psi) \rangle$ 
\State $T_0$\,$\leftarrow$\,$\norm{|P_0\rangle}$
\State $k$\,$\leftarrow$\,$1$
\Repeat
    \State $|P_{k-1}\rangle$\,$\leftarrow$\, $|P_{k-1}\rangle / T_{k-1}$ 
    \State $W_k$\,$\leftarrow$\,$\mathrm{diag}(|P_{k-1} \rangle)$ 
    \State $|P_k\rangle$\,$\leftarrow$\,$W_k|P_{k-1} \rangle$ 
    \State $T_k$\,$\leftarrow$\,$\norm{|P_{k} \rangle}$
    \State $k$\,$\leftarrow$\,$k+1$
\Until{$|1-T_k/T_{k-1}|$\,$\leq$\,$\epsilon$}
\State $\nu$\,$\leftarrow$\,$N+2\log_2 T_k$.
\end{algorithmic}
\end{algorithm}

The information about the stabilizer group of $| \psi  \rangle$ can be extracted from $| G(\psi)  \rangle$, since we have
\begin{equation}
    \langle \boldsymbol{\alpha} | G(\psi)  \rangle = \begin{cases}
        \sqrt{2^{\nu-N}}, & \text{if $P_{\boldsymbol{\alpha}} | \psi  \rangle = \pm | \psi  \rangle$} \\
        0, & \text{otherwise }.
    \end{cases}
\end{equation}
The unsigned generators of the stabilizer group can be extracted using perfect MPS sampling~\cite{Ferris2012} on $| G(\psi)  \rangle$. The protocol is equivalent to learning a stabilizer state by Bell sampling~\cite{montanaro2017learning}, which can be done efficiently in $O\left(N^3\right)$ time. Once all the unsigned generators are found, the signs of the generators can be extracted from $|P(\psi) \rangle$. In this way, the generators of the stabilizer group can be determined in $O\left(N^2 (N-\nu) + N(N-\nu) \chi_G^2\right)$ time, where $\chi_G$ is the bond dimension of $| G(\psi)  \rangle$.

\bibliographystyle{apsrev4-1}
\bibliography{bibliography}

\clearpage
\onecolumngrid
\section{Supplemental Material}

We present additional information on (1) short review on Clifford circuits, (2) the properties of the stabilizer entropy, stabilizer nullity, and Bell magic, (3) the convergence analysis of the nullity algorithm, (4) the procedure to learn states prepared with few non-Clifford gates, (5) generalization to MPO, (6) generalization to qudits, (7) the MPS compression, (8) possibility for doing transverse contractions, and (9) additional numerical results.

\renewcommand{\theequation}{S\arabic{equation}}
\renewcommand{\thefigure}{S\arabic{figure}}
\setcounter{equation}{0}
\setcounter{figure}{0}

\newcommand{\mainref}[2]{#1}

\input{./supmat}

\end{document}

%% file: supmat.tex
\setcounter{equation}{0}
\setcounter{figure}{0}
\setcounter{table}{0}
\setcounter{page}{1}
\makeatletter
\renewcommand{\theequation}{S\arabic{equation}}
\renewcommand{\thefigure}{S\arabic{figure}}
\renewcommand{\thetable}{S\arabic{table}}
\renewcommand{\bibnumfmt}[1]{[S#1]}

\subsection{Clifford circuits}

The Clifford group on $N$ qubits
is defined as the normalizer of the $N$-qubit Pauli group, and can be generated by the Hadamard gate, the $\pi/4$-phase gate, and the controlled-NOT gate, defined in the following
\be 
H= \frac{1}{\sqrt{2}} \begin{pmatrix}
1 & 1\\
1 & -1
\end{pmatrix}
\ee

\be 
S=\begin{pmatrix}
1 & 0\\
0 & i
\end{pmatrix}
\ee

\be 
CNOT= \begin{pmatrix}
1 & 0 & 0 & 0\\
0 & 1 & 0 & 0\\
0 & 0 & 0 & 1\\
0 & 0 & 1 & 0
\end{pmatrix}
\ee
, respectively.

The celebrated Gottesman-Knill theorem states that any quantum computation with only Clifford gates can be efficiently simulated by a classical computer, despite being capable of generating high amount of entanglement and exhibiting very rich structures. In other words, nonstabilizer features are needed in order to enable universal quantum computation and achieve the desired quantum computational advantages.
A suitable extension of the Clifford circuits allows to perform universal quantum computation, as for instance, adding the $T$ gate  
\be 
T=\begin{pmatrix}
1 & 0\\
0 & e^{i\pi/4}
\end{pmatrix}
\ee
, which, combined with the Clifford gates, creates a universal gate set. Another commonly used nonstabilizer gate is the $CCZ$ (controlled-controlled-$Z$) gate
\be 
CCZ = \mathrm{diag}(1,1,1,1,1,1,1,-1).
\ee

\subsection{Measures of nonstabilizerness}
\subsubsection{Stabilizer Rényi entropy}
In this section, we define the stabilizer R\'enyi entropy and we briefly state some of its key properties to allow easy access to the main results of the paper.

Consider the $d=2^N-$dimensional Hilbert space of $N$ qubits $\mathcal H\simeq \mathbb C^{\otimes 2N}$.
Let us call $\mathcal{P}_N$ the group of all $N$-qubit Pauli operators with phase $1$, and define $\Xi_{\psi} (P)=d^{-1} \tr(P \Psi)=\langle P \rangle_{\psi}$ as the squared (normalized) expectation value of $P$ in the pure state $|\psi\rangle$ with density matrix $\Psi=|\psi\rangle \langle \psi|$. Moreover, $\Xi_{\psi}$ is the probability of finding $P$ in the representation of the state $|\psi\rangle$. 

The SREs are defined in the Eq. \eqref{eq:SRE}. For three common choices of $n$ the stabilizer R\'enyi entropy reads
\begin{equation}
	M_{n}(|\psi\rangle) =
	\begin{cases}
		\log_2 \left( \lvert \left \lbrace  P \in \mathcal{P}_N :  \langle P \rangle_\psi \neq  0  \right \rbrace \lvert \right)-N & n \rightarrow 0 \\
		-\sum_P 2^{-N} \langle P \rangle_\psi^2 \log_2 \left( \langle P \rangle_\psi^2 \right) & n \rightarrow 1 \\
		-\log_2 \left(\sum_P 2^{-N} \langle P \rangle_\psi^4   \right) & n =2
	\end{cases}
\end{equation}
where $P\in\mathcal{P}_N$ is an element of the group of all $N$--qubit Pauli strings with +1 phases.
We list some key properties of the stabilizer $\alpha$--R\'enyi entropies, alongside the references that contain the respective proofs:
\begin{enumerate}[] 
    \item \emph{Faithfulness}: $M_n \left(|\psi\rangle \right)=0$ if and only if $|\psi \rangle$ is a stabilizer state (see Ref.~\cite{leone2022stabilizer}).
    \item \emph{Stability under free operations}: For any unitary Clifford operator $C$ and state $|\psi\rangle$ it holds that \mbox{$M_n(C |\psi \rangle) = M_\alpha\left( |\psi \rangle \right)$} (see Ref.~\cite{leone2022stabilizer}).
    \item \emph{Additivity}: $M_n \left( |\psi \rangle \otimes |\phi \rangle \right) = M_n \left(|\psi \rangle \right)+ M_n \left(|\phi \rangle \right)$ (see Ref.~\cite{leone2022stabilizer}).
    \item \emph{Bounded}: For any $N$-qubit state $|\psi\rangle$ it holds that $0\leq M_n(|\psi\rangle)<N$ (see Ref.~\cite{leone2022stabilizer}).
    \item  $M_{n^{\prime}}(|\psi\rangle)<M_n(|\psi\rangle)$ for $n^{\prime}>n$ (see Ref.~\cite{haug2023stabilizer}).
    \item The stabilizer entropies consitute a lower bound to the so--called $T$-count $t(|\psi\rangle)$ of a state: \mbox{$M_n\left( |\psi \rangle \right)< t(|\psi\rangle)$} (see Ref.~\cite{gu_little_2023}).
    \item For $\alpha > 1/2$ the stabilizer entropies constitute  a lower bound to the so--called ``robustness of magic'': $M_n\left( |\psi \rangle \right)<\mathcal{R}_{\psi}$ where $\mathcal{R}_{\psi}= \min_{x} \left \lbrace ||x||_1 | |\psi \rangle \langle \psi|=\sum_i x_i \sigma_i, \sigma_i \in STAB \right \rbrace$ (see Refs.~\cite{heinrich2019robustness,leone2022stabilizer}).
\end{enumerate}

\subsubsection{Bell magic}
\begin{figure}
        \centering
        \includegraphics[width=1.0\textwidth]{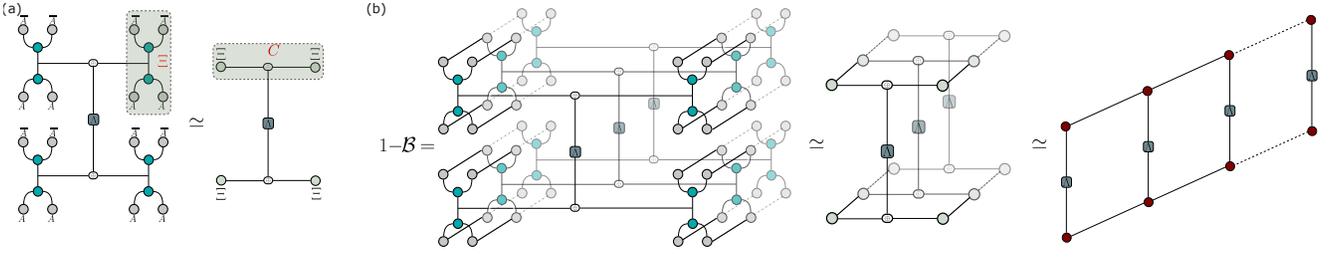}
        \caption{Tensor network representation of the additive Bell magic.}.
        \label{fig:bell_magic_tn}
    \end{figure}

In this section, we discuss the properties of Bell magic. Its definition is 
\be
\mathcal{B}=\sum \Xi(\overline{r}) \Xi(\overline{r}^{\prime}) \Xi(\overline{q}) \Xi(\overline{q}^{\prime}) || [\sigma_{\overline{r} \oplus \overline{r}^{\prime}},\sigma_{\overline{q} \oplus \overline{q}^{\prime}}] ||_{\infty} ,
\ee
where $\Xi(\overline{r})$ is the probability of the outcome $\overline{r}$ if we perform the Bell measurement on two copies of pure state $|\psi \rangle \otimes |\psi \rangle$
\be 
\Xi(\overline{r})= \langle \psi | \langle \psi | O_{\overline{r}} |\psi \rangle |\psi \rangle = 2^{-N} |\langle \psi | \sigma_{\overline{r}} |\psi^{\star} \rangle|^2,
\ee
with $O_{\overline{r}}= |\sigma_r \rangle \langle \sigma_r| $ is a projector onto a product of Bell states and $|\psi^{\star} \rangle$ denotes the complex conjugate of $|\psi \rangle$. 
The infinity norm is zero when the Pauli strings commute. As a measure of magic, $\mathcal{B}=0$ only for pure stabilizer states $|\psi_{\rm STAB} \rangle$ and  $\mathcal{B}>0$ otherwise.
$\mathcal{B}$ is also invariant under Clifford circuits
$U_C$ that map stabilizers to stabilizers for example $\mathcal{B}(U_C |\psi \rangle)$. Moreover, Bell magic is constant under composition with any stabilizer state, i.e. if $|\psi_{\rm STAB} \rangle$ then $\mathcal{B}(|\psi \rangle \otimes |\psi_{\rm STAB} \rangle )= \mathcal{B}(|\psi \rangle)$.
We further define the additive Bell magic:
\begin{equation}
\mathcal{B}_a = -\log_2 \left( 1- \mathcal{B}\right) .
\end{equation}
$\mathcal{B}_a$ has the same properties of $\mathcal{B}$ and, further, it is also additive
\begin{equation}
    \mathcal{B}_a (|\psi \rangle \otimes |\phi \rangle) = \mathcal{B}_a(|\psi \rangle)+\mathcal{B}_a(|\phi \rangle).
\end{equation}
Moreover, $\mathcal{B}_a$  has the operational meaning
as the number of initial magic states $|T\rangle$. For example, if we consider the state $|\psi \rangle= |T\rangle^{\otimes k} \otimes |0\rangle^{\otimes N-k}$ consisting of a tensor product of $k$
magic states and otherwise the stabilizer state $|0\rangle$, then the additive Bell magic is 
\begin{equation}
    \mathcal{B}_a \left( |T\rangle^{\otimes k} \otimes |0\rangle^{\otimes N-k} \right) = k .
\end{equation}

\subsubsection{Stabilizer nullity}
In this section, we introduce the stabilizer nullity, a function $\nu(|\psi\rangle)$ of any pure state $|\psi\rangle$ that is non-increasing under stabilizer operations. The stabilizer nullity is surprisingly powerful given its simplicity: it is the number of qubits that $|\psi \rangle$ is hosted in, minus the number of
independent Pauli operators that stabilize $|\psi \rangle$. 

Before introducing the definition of nullity, let us first recall the definition of a stabilizer state and introduce a slight generalization of it. 
Let $|\psi \rangle$ be a non-zero n-qubit state. The stabilizer of $|\psi \rangle$, denoted ${\rm Stab}(|\psi \rangle)$, is the sub-group of the Pauli group $P_N$ on $N$ qubits for which $|\psi \rangle$ is a $+1$ eigenstate, that is ${\rm Stab}(|\psi \rangle) = \{P \in P_N : P|\psi \rangle = |\psi \rangle\}$. The states for which the size of the stabilizer is $2^N$ are called stabilizer states. States for which the stabilizer contains only the identity matrix are said to have a trivial stabilizer. If Pauli $P$ is in ${\rm Stab}(|\psi \rangle)$, we say that $P$ stabilizes $|\psi \rangle$. Now we can define the  stabilizer nullity as 
\be 
\nu \left(|\psi \rangle \right)=N-\log\left( |{\rm Stab}(\psi)| \right) .
\ee

Moreover, one of the most important property of ${\rm Stab}(\psi)$  is that let $P$ be an $N$-qubit Pauli matrix
and suppose that the probability of a $+1$ outcome when measuring $P$ on $|\psi\rangle$ is non-zero. Then
there are two alternatives for the state $|\phi\rangle$ after the measurement: either ${\rm Stab}(|\phi \rangle)={\rm Stab}(|\psi \rangle)$,
or ${\rm Stab}(|\phi \rangle) \geq 2 {\rm Stab}(|\psi \rangle)$, both of which satisfy $\nu(|\phi \rangle)<\nu(|\psi\rangle)$.
Following this previous property of ${\rm Stab}(|\psi \rangle)$ , it is easy to demonstate that the stabilizer nullity $\nu$ is invariant under Clifford unitaries, is non-increasing under Pauli measurements, and
is additive under the tensor product. Moreover, as $\nu=0$  when $|\psi \rangle$ is a stabilizer state,
the stabilizer nullity is invariant under the inclusion or removal of stabilizer states.

\subsubsection{Magic gap}
Magic gap is defined as \cite{Bu2023}

\begin{equation}
    MG(| \psi \rangle) = 1- \max_{\boldsymbol{\alpha}, \langle \psi | P_{\boldsymbol{\alpha}} | \psi \rangle \neq \pm1} |\langle \psi | P_{\boldsymbol{\alpha}} | \psi \rangle|.
\end{equation}
It controls the rate of convergence of Algorithm \ref{alg:nullity} to compute the nullity, as explained in the next section. Additionally, it can be calculated using similar approach as the nullity. To do so, we first project out the Pauli strings in the stabilizer group:  $| P'(\psi) \rangle =  |P(\psi) \rangle - \sqrt{2^{N-\nu}}W_{\infty}|P(\psi) \rangle$ . Applying Algorithm \ref{alg:nullity} to $| P'(\psi) \rangle$, we will obtain the fixed point $|G'(\psi) \rangle$ that encodes the set of Pauli strings with the second largest values (in magnitude) in the Pauli spectrum of $| \psi \rangle$. Finally, using perfect MPS sampling \cite{Ferris2012} on $|G'(\psi) \rangle$, we can extract a specific Pauli string whose expectation value yields the magic gap.

\subsection{Convergence of the nullity algorithm}
The rate of convergence of Algorithm \ref{alg:nullity} is determined by the magic gap, through the following upper bound \cite{bu2024entropic}
\begin{equation} \label{eq:upper_bound}
    |\nu_k - \nu|  \leq 3\log_2 \left[1 + (1-MG(| \psi \rangle)^{2^{k+1}-2} 2^\nu  )\right]. 
\end{equation}
Here, $\nu_k$ is the approximate nullity obtained if the algorithm is terminated at the $k$-th iteration. One can verify that
\be 
\nu_k = \log_2 \left \lbrace \sum_{\boldsymbol{\alpha}} \frac{|\langle \psi | P_{\boldsymbol{\alpha}} | \psi \rangle|^{2^{k+1}}}{2^N} \right \rbrace - 2 \log_2 \left \lbrace \sum_{\boldsymbol{\alpha}} \frac{|\langle \psi | P_{\boldsymbol{\alpha}} | \psi \rangle|^{2^{k}}}{2^N} \right \rbrace  \,  . 
\ee
 Based on the upper bound in Eq.~\eqref{eq:upper_bound}, approximating the nullity with a fixed error requires $k=O\left(\log_2\frac{\nu}{-\log_2\left[1-MG(| \psi \rangle)\right]}\right)$ iterations. It can also be shown that $\nu_k$ monotonically increases with $k$:
 \begin{equation}
 \begin{split}
     \nu_{k+1} - \nu_k &= \left( \log_2 \left \lbrace \sum_{\boldsymbol{\alpha}} \frac{|\langle \psi | P_{\boldsymbol{\alpha}} | \psi \rangle|^{2^{k+2}}}{2^N} \right \rbrace - 2 \log_2 \left \lbrace \sum_{\boldsymbol{\alpha}} \frac{|\langle \psi | P_{\boldsymbol{\alpha}} | \psi \rangle|^{2^{k+1}}}{2^N} \right \rbrace \right) \\
     & \quad \quad - \left( \log_2 \left \lbrace \sum_{\boldsymbol{\alpha}} \frac{|\langle \psi | P_{\boldsymbol{\alpha}} | \psi \rangle|^{2^{k+1}}}{2^N} \right \rbrace - 2 \log_2 \left \lbrace \sum_{\boldsymbol{\alpha}} \frac{|\langle \psi | P_{\boldsymbol{\alpha}} | \psi \rangle|^{2^{k}}}{2^N} \right \rbrace \right) \\
     &= \log_2 \left \lbrace \sum_{\boldsymbol{\alpha}} \frac{|\langle \psi | P_{\boldsymbol{\alpha}} | \psi \rangle|^{2^{k+2}}}{2^N} \right \rbrace + 2 \log_2 \left \lbrace \sum_{\boldsymbol{\alpha}} \frac{|\langle \psi | P_{\boldsymbol{\alpha}} | \psi \rangle|^{2^{k}}}{2^N} \right \rbrace - 3\log_2 \left \lbrace \sum_{\boldsymbol{\alpha}} \frac{|\langle \psi | P_{\boldsymbol{\alpha}} | \psi \rangle|^{2^{k+1}}}{2^N} \right \rbrace \\
     &= \log_2 \left \lbrace \frac {\left( \sum_{\boldsymbol{\alpha}} |\langle \psi | P_{\boldsymbol{\alpha}} | \psi \rangle|^{2^{k+2}} \right) \left( \sum_{\boldsymbol{\alpha}} |\langle \psi | P_{\boldsymbol{\alpha}} | \psi \rangle|^{2^{k}} \right)^2}{\left( \sum_{\boldsymbol{\alpha}} |\langle \psi | P_{\boldsymbol{\alpha}} | \psi \rangle|^{2^{k+1}} \right)^3} \right \rbrace \\
     &\geq 0.
\end{split}
 \end{equation}
The last line follows by applying H\"older's inequality to the term inside the logarithm in the second-to-last line. Consequently, $\nu_k$ provides a rigorous lower bound to the nullity for each $k$.

\subsection{Learning states prepared with few non-Clifford gates}
In this section, we discuss how to learn the full description of states prepared with few non-Clifford gates, i.e., states with small $\nu$, within our MPS framework. Let $| \psi \rangle$ be a pure state of $N$ qubits whose stabilizer group is generated by $m$ Pauli strings $s_j P_{\boldsymbol{\alpha}_j}$ for $j=1,2,\cdots,m$ and $s_j = \pm 1$. We will make use of the algebraic structure of $T$-doped stabilizer states \cite{leone2023learning}:
\begin{equation}
    |\psi \rangle \langle \psi | = \frac{1}{2^N} \sum_{i=0}^l \langle \psi | P_{\boldsymbol{\beta}_i} | \psi \rangle P_{\boldsymbol{\beta}_i} \prod_{j=1}^m (I + s_j P_{\boldsymbol{\alpha}_j}),
\end{equation}
where $P_{\boldsymbol{\beta}_i}$ for $i=0,1, \cdots,l$ are referred to as the bad generators, and $P_{\boldsymbol{\beta}_0}=I$. Therefore, to fully characterize the state $| \psi \rangle$, it is sufficient to learn its stabilizer group and the bad generators. In the main text, we have detailed how to learn the stabilizer group from $|G(\psi) \rangle$ obtained from Algorithm \ref{alg:nullity}. The next step is to learn the $l$ bad generators. One possible approach involves constructing a Clifford circuit $C$ such that $C | \psi \rangle = | \phi \rangle \otimes | x \rangle$, where $|\phi \rangle$ is a state of $\nu$ qubits and $| x \rangle$ is a computational basis state of $N-\nu$ qubits \cite{leone2023learning-2}. The bad generators can then be learned directly from the Pauli vector of $|\phi \rangle$. However, applying a Clifford ciruit to an MPS in general leads to a significant increase in the bond dimension of the MPS. It is thus preferable to perform the task in a way that avoids explicit application of a Clifford circuit.

We achieve this by iteratively extracting the bad generators with large expectation values. First, we project out the stabilizer group from the Pauli vector: $| P'(\psi) \rangle =  |P(\psi) \rangle - \sqrt{2^{N-\nu}}W_{\infty}|P(\psi) \rangle$. Then, we repeat Algorithm \ref{alg:nullity} to obtain a fixed point $| G^{(2)} (\psi) \rangle $ encoding a union of left disjoint cosets $P_{\boldsymbol{\beta}_1} {\rm Stab}(\psi ) \cup \cdots \cup P_{\boldsymbol{\beta}_{l_2}} {\rm Stab}(\psi)$. The Pauli strings $P_{\boldsymbol{\beta}_{1}}, \cdots,P_{\boldsymbol{\beta}_{l_2}}$ are those with the second-largest expectation values (in magnitude) in the Pauli spectrum of $| \psi \rangle$. This process repeats, each time projecting out previously found bad generators and using Algorithm \ref{alg:nullity} to identify the next cosets of Pauli strings with large expectation values. The procedure stops when the Pauli vector is finally empty. In the end, we will have a set of fixed points $| G^{(n)} (\psi) \rangle$, each encoding the bad generators with the $n$-th largest expectation values (in magnitude) in the Pauli spectrum of $| \psi \rangle$. Then, similarly as the stabilizer generators, the bad generators can be learned using perfect MPS sampling on each $| G^{(n)}(\psi)  \rangle$. The total time complexity to obtain the fixed points is $O\left(l N \log(\nu)\chi_{\mathrm{max}}^4\right)$. Importantly, this allows us to directly access the full content of the Pauli spectrum. Learning the generators themselves then takes $O\left(l \left[N^2(N-\nu) +  N(N-\nu) \chi_G^2\right]\right)$. Since $l<4^\nu$, our algorithm learns the generators of the state efficiently for $\nu=O(\log N)$.

\subsection{Generalization to matrix product operators}
The technique presented in the main text can be straightforwardly adapted to matrix product operators (MPO), which represent mixed states. We consider a density matrix $O$ of $N$ qubits represented in the following MPO form:
\begin{equation}
    O =\sum_{\boldsymbol{s},\boldsymbol{s'}} U^{s_1,s'_1}_1 U^{s_2,s'_2}_2 \cdots U^{s_N,s'_N}_N | s_1, \cdots, s_N \rangle \langle s'_1, \cdots, s'_N |
\end{equation}
with $U^{s_i,s'_i}_i$ being $\chi \times \chi$ matrices, except at the left (right)
boundary where $U^{s_1,s'_1}$ (or $U^{s_N,s'_N}$) is a $1 \times \chi$ ($\chi \times 1$) row (column) vector.

The Pauli vector $|P(O) \rangle$ can be obtained in a similar way as in MPS, namely
\begin{equation}
|P(O) \rangle=\sum_{\boldsymbol{\alpha}} V^{\alpha_1}_1 V^{\alpha_2}_2 \cdots V^{\alpha_N}_N | \alpha_1, \cdots, \alpha_N \rangle
\end{equation}
where $V^{\alpha_i}_i = \sum_{a,b} \langle a | P_{\alpha_i} | b \rangle U^{a,b}_i /  \sqrt{2}$ are $\chi \times \chi$ matrices. The procedure above can be seen as MPO version of the method recently discussed in Ref. \cite{hamaguchi2023handbook} to obtain Pauli vector representation from the full density matrix. Notice that, unlike in the MPS case, in this case the bond dimension remains the same. Indeed, the transformation above is simply a local basis transformation from the computational basis to the Pauli basis. Note also that the norm of $|P(O) \rangle$ is $\Tr\left[O^2\right]$, which is generally different from 1.  Using $|P(O) \rangle$, one can compute the SRE, nullity and the Bell magic of $O$ in the same way as in the MPS case (see Main text). However, we note that these measures of nonstabilizerness are only faithful for pure states. Nevertheless, we expect that this technique could be useful, e.g., to compute the mana~\cite{veitch2012negative,veitch2014resource,wang2019quantifying}, which is a good nonstabilizerness measure for mixed states.

\subsection{Generalization to qudits}
The generalization to $d-$state qudits is straightforward, by considering the $d^2$ generalized Pauli operators defined for qudits. With this, one can gain access to the qudit SRE for integer $n>1$. One key difference with the qubit case is that the Pauli operators are not Hermitian for $d>2$, and thus the Pauli vector is not necessarily real in the Pauli basis. 

For odd prime $d$, one can also consider the set of phase-space operators, defined as
\begin{equation}
    A_0 = \frac{1}{d^N} \sum_\textbf{u} P_{\textbf{u}}, \quad A_{\textbf{u}} = P_{\textbf{u}}A_0P_{\textbf{u}}^\dagger,
\end{equation}
which provide an orthonormal basis for Hermitian operators in $\mathbb{C}^{d^N \otimes d^N}$. In analogy to the Pauli vector, one can compute the vector storing the discrete Wigner function 
\begin{equation} \label{eq:wigner}
    W_{\rho}(\mathbf{u}) = \frac{1}{d^N} \Tr(A_{\mathbf{u}}\rho).
\end{equation}
 One can then compute the mana entropy \cite{tarabunga2023critical} for integer $n>1$ with similar technique as the SRE. The mana itself, which corresponds to $n=1/2$, is not accessible with the replica method. Nevertheless, one can perform sampling on the MPS containing the discrete Wigner function to compute the mana.

\subsection{MPS compression}
As mentioned in the main text, the MPS $| P^{(n)}(\psi) \rangle$ should be compressed to keep the cost manageable. There are a few methods to perform the compression, such as the density matrix algorithm~\cite{McCulloch2007}, the SVD compression \cite{Stoudenmire2010}, and variational compression~\cite{Schollwck2011}. We refer to Refs.~\cite{Schollwck2011,McCulloch2007} for details on the compression methods.

To compress the Pauli vector $| P(\psi) \rangle$, we perform the SVD compression by iteratively truncating the bond dimension to $\chi_P$ from left to right, while moving the orthonormality center. We recall that $| P(\psi) \rangle$ without compression is automatically orthonormalized, which implies that the compression is (globally) optimal. The overall cost of the compression is $O\left( \chi_P^2 \chi^2 + \chi^3 \chi_P \right)$. Notice however that the cost of exact construction (without performing truncation) is $O\left(\chi^4\right)$, which results in the Pauli vector $| P(\psi) \rangle$ of bond dimension $\chi_P=\chi^2$ that is already in a canonical form. Therefore, if $\chi_P$ cannot be set much lower than $\chi^2$ without losing too much accuracy, exact construction offers a more efficient alternative.

Similar SVD compression can be performed to compress the MPO-MPS multiplication $W | P(\psi) \rangle$. However, the resulting MPS is no longer orthonormalized, and the considerations above do not apply. Nevertheless, as argued in Refs.~\cite{Schollwck2011,Stoudenmire2010} , the SVD compression in the MPO-MPS product would still yield a good result, particularly if both the MPO and MPS are orthonormalized (which is true in our case). 

In our computations of the SRE and the Bell magic, we have performed the compression using only the SVD compression. We have checked with bond dimension up to $\chi_P=100$ that the results using the SVD compression is consistent with the solution obtained by the density matrix algorithm, which is optimal but more costly.

To calculate the stabilizer nullity, we find that the density matrix algorithm is more reliable to obtain the correct result. Therefore, we used the density matrix algorithm to obtain the stabilizer nullity of the Ising and XXZ chain. However, the density matrix algorithm is too costly for the simulation of random Clifford circuits. In that case, we instead perform only SVD compression.

\subsection{Transverse contraction}
An alternative way to perform the contraction of the two-dimensional tensor network in Fig.~\ref{fig:folding_mps}  (c) is by contracting the tensors in the transversal (space) direction \cite{PhysRevLett.102.240603,muller2012tensor,hastings2015connecting}. To do so, we first contract the $2n$ tensors in the first site to form a transfer matrix with $2n$ indices, each with bond dimension $\chi^2$. Then, we iteratively absorb the tensors on the right to the transfer matrix, up until the rightmost tensors. Without compression, the cost of this contraction scheme is $O\left(\chi^{4n+2}\right)$, which is cheaper than the exact contraction in the direction of Rényi index, or the contraction in Ref.~\cite{haug2023quantifying}. Of course, the contractions can also be done approximately by representing the transfer matrix as an MPS. Whether or not this would yield a better performance compared to the approximate contraction in the direction of Rényi index is an intriguing question that we leave for future research avenue.

In the case of translation-invariant (TI) MPS in the thermodynamic limit, we can compute the SRE by introducing the transfer matrix
\begin{equation}
\begin{split}
    \tau &= \sum_\alpha B^{(n)\alpha} \otimes B^{(n)\alpha}     \\
        &= \sum_\alpha   (B^{\alpha})^{\otimes 2n}.  
\end{split}
\end{equation}
Here, we recall that $B^{(n)\alpha}$ is the local tensor of $|P^{(n)}(\psi) \rangle$, which is site independent for TI MPS. The transfer matrix $\tau$ is identical to the one introduced in Ref.~\cite{haug2023quantifying}, however the local tensors that build $\tau$ differ. In particular, with our approach, the transfer matrix can be viewed as an MPO with physical dimension $\chi^2$ and constant bond dimension of 4, i.e., the MPO satisfies an area law. The calculation of the SRE is then reduced to the computation of the dominant eigenvalue of $\tau$. This can be done by approximating the dominant eigenvector $|L \rangle$ as an MPS, and performing power iteration or Lanczos algorithm by repeated MPO-MPS multiplication.

\subsection{Additional numerical results}

\subsubsection{Stabilizer Rényi entropies}
\begin{figure}
        \centering
        \includegraphics[width=0.7\textwidth]{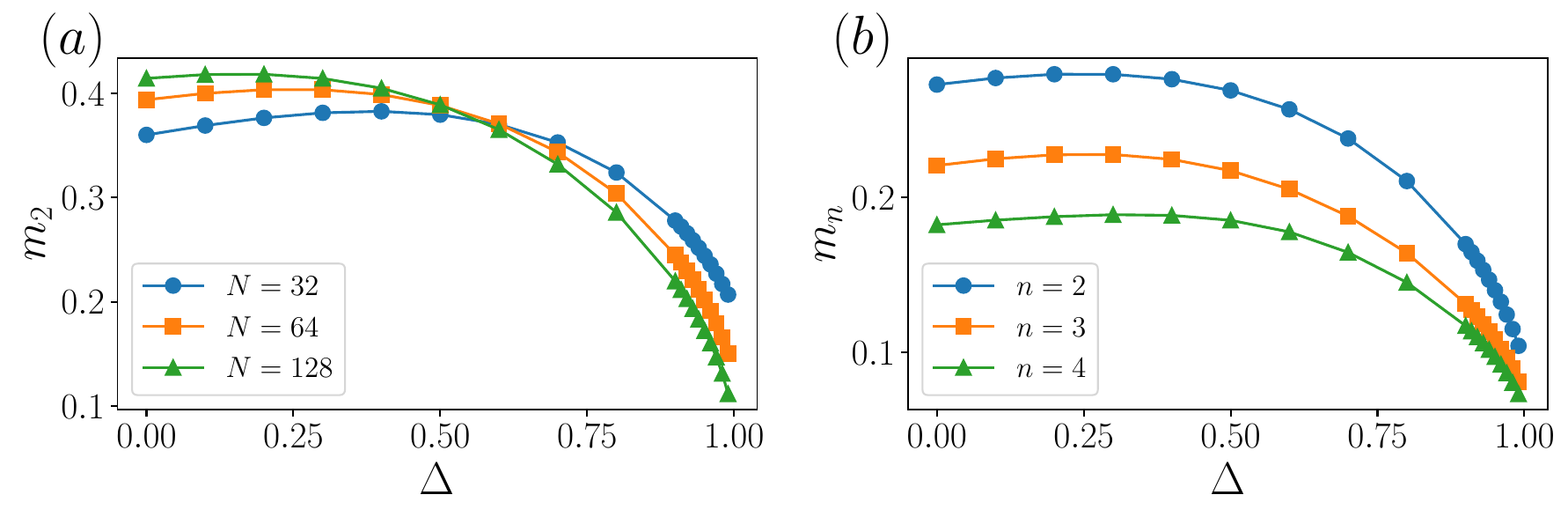}
        \caption{SRE density $m_n=M_n/N$ for the ground state of the XXZ chain as a function of the anisotropy $\Delta$ for (a) $n=2$ in various system sizes and (b) for $n\in \{2,3,4\}$ with $N=64$.}.
        \label{fig:replica}
    \end{figure}

We benchmarked the calculations of the SRE in the ground states of the XXZ chains, 
\begin{equation}
H_{\text{XXZ}}=-\sum_{\langle i,j \rangle} \left[ \sigma^x_i \sigma^x_j + \sigma^y_i \sigma^y_j + \Delta \sigma^z_i \sigma^z_j \right].
\end{equation} We first obtain the ground states using DMRG with $\chi=60$ and compress the bond dimension of the Pauli vector to $\chi_P=400$. Fig.~\ref{fig:replica} (a) shows the results for $n=2$ in various system sizes up to $N=128$. We note that the Rényi-2 SRE could not be computed accurately for $N>30$ in the previous study~\cite{haug2023stabilizer}. The discussion about convergence with bond dimension within our approach can be found in the next section. Moreover, our method enables easier access to higher index SREs, as shown in Fig.~\ref{fig:replica} (b) for $n\in \{2,3,4\}$. The ability to compute higher index SREs offers significant advantages. First, it allows us to differentiate between typical and atypical quantum states based on how their SREs change with the Rényi index \cite{turkeshi2023measuring}. Second, as detailed in the main text, higher index SREs pave the way for calculating the stabilizer nullity.

\subsubsection{Convergence with bond dimension in replica MPS}
\begin{figure}
        \centering
        \begin{overpic}[width=0.35\linewidth]{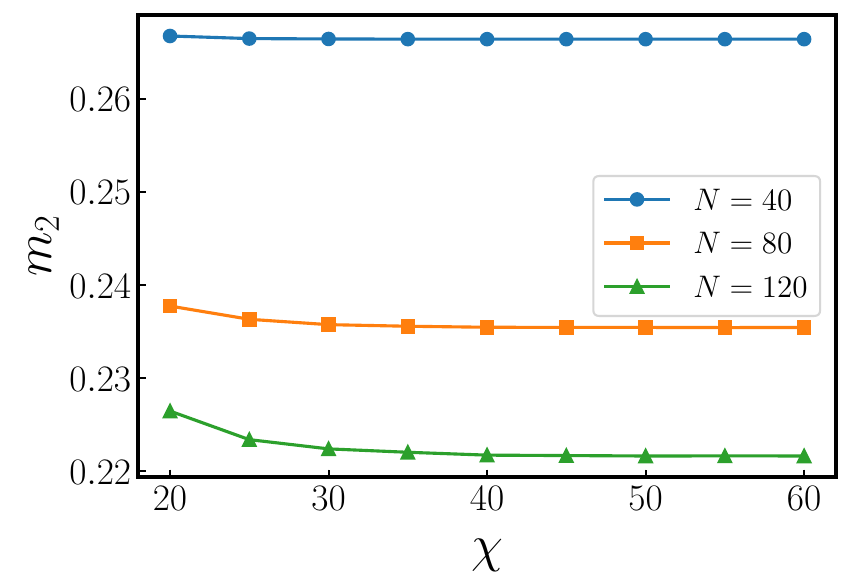}
        \put (-1,70) {{\textbf{(a)}}}
        \end{overpic}
    \begin{overpic}[width=0.35\linewidth]{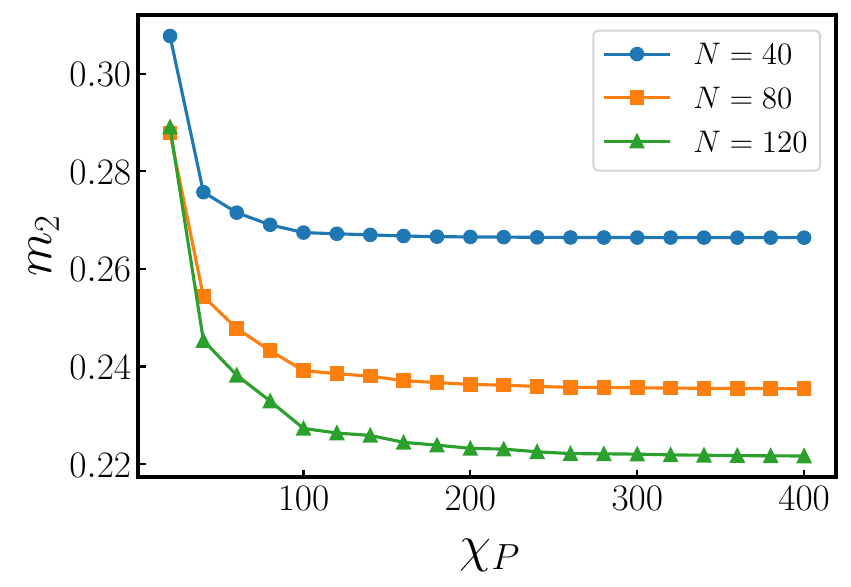}
    \put (-1,70) {{\textbf{(b)}}}
    \end{overpic}
    \begin{overpic}[width=0.35\linewidth]{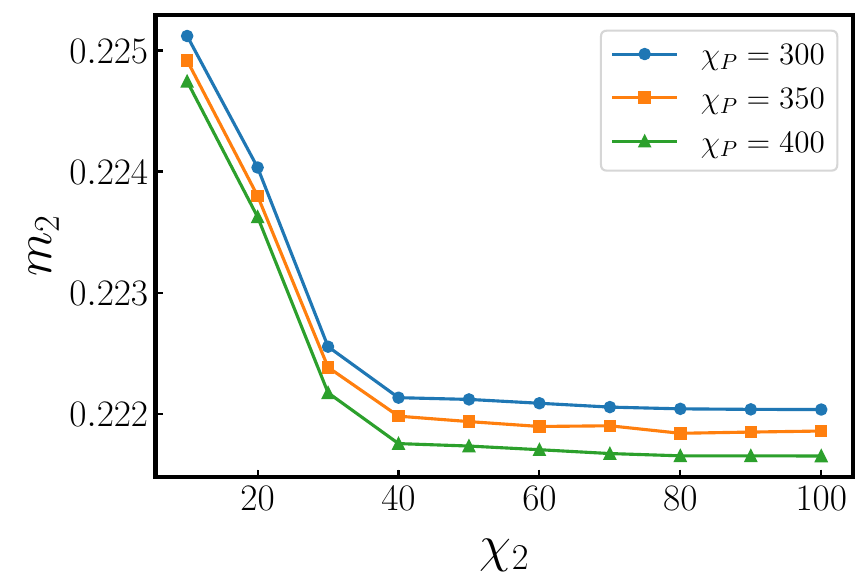}
    \put (-1,70) {{\textbf{(c)}}}
    \end{overpic}
    \begin{overpic}[width=0.35\linewidth]{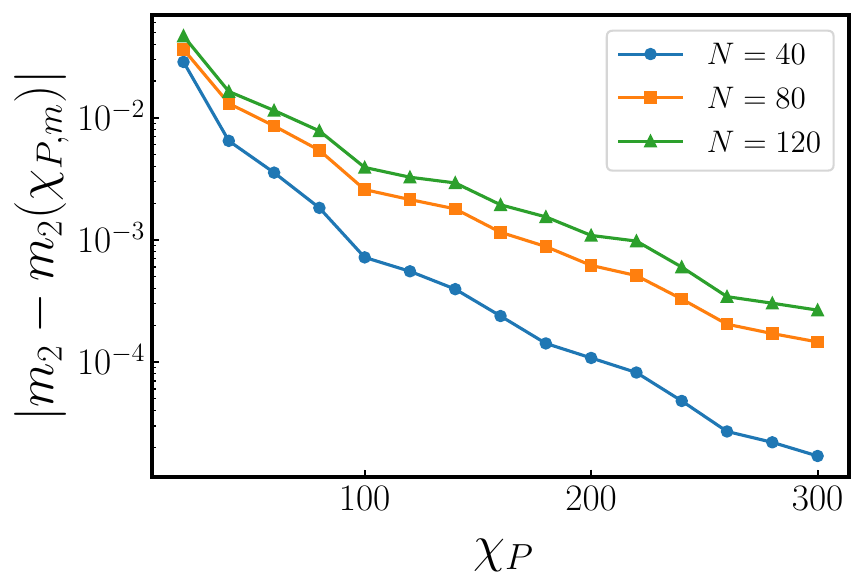}
    \put (-1,70) {{\textbf{(d)}}}
    \end{overpic}
    
        \caption{SRE density $m_2=M_2/N$ for the ground states of the XXZ chain with anisotropy $\Delta=0.9$ (a) as a function of bond dimension $\chi$ with fixed $\chi_P=400$ and $\chi_2=100$, (b) as a function of bond dimension $\chi_P$ with fixed $\chi=60$ and $\chi_2=100$, and (c) as a function of $\chi_2$ with fixed $\chi_P \in \{ 300,350,400\}$ and $\chi=60$ for $N=120$. (d) Difference of $m_2$ computed for bond dimension $\chi_P$ and the maximum bond dimension $\chi_{P,m} = 400$ at fixed $\chi_2=100$. }.
        \label{fig:SRE_replica}
    \end{figure}

In our simulations, we have studied the accuracy of our approach by checking the convergence of our results with bond dimension. 
Fig. \ref{fig:SRE_replica} illustrates an example of the dependence of the SRE $m_2$ in the ground states of the XXZ chain. In particular, we studied the effect of increasing $\chi$, $\chi_P$ and $\chi_2$. We see that as the bond dimensions are increased, the SRE eventually converges to a constant. Interestingly, we find that $\chi_2$ can be set to a smaller value than $\chi_P$. Moreover, we see that the results are not accurate for $\chi=12$, which is the maximum achievable with the exact replica method (see Refs. \cite{haug2023quantifying,haug2023stabilizer}). In Fig. \ref{fig:SRE_replica} (d), we show that the decay of the error as a function of the bond dimension appears to be exponential, consistent with the findings of Ref. \cite{haug2023quantifying}.   

\subsubsection{Bell magic}
The tensor network representation of $1-\mathcal{B}$, where $\mathcal{B}$ is the Bell magic, is shown in Fig. \ref{fig:bell_magic_tn}. The additive Bell magic is then given by $\mathcal{B}_a = -\log(1-\mathcal{B})$ (see Eq. \eqref{eq:additive_bell_magic} in the main text).

\begin{figure}
        \centering
        \includegraphics[width=0.4\textwidth]{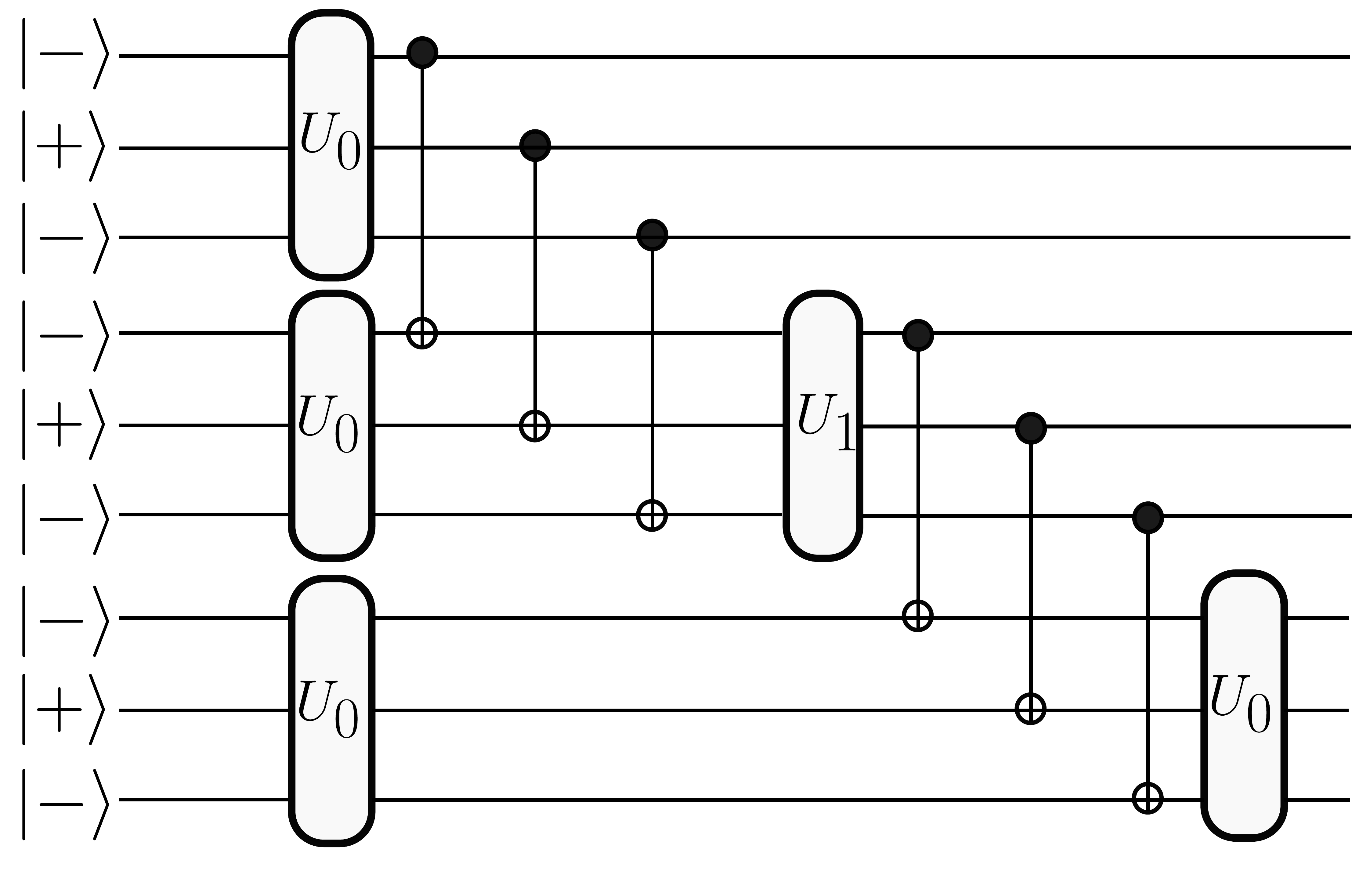}
        \caption{The scrambling circuit recently experimentally realized in Ref. \cite{Bluvstein2023} to measure the additive Bell magic for $N=9$. The gates $U_0$ and $U_1$ are as defined in Ref. \cite{Bluvstein2023}.}.
        \label{fig:circuit_L9}
    \end{figure}
\begin{figure}
        \centering
        \begin{overpic}[width=0.70\linewidth]{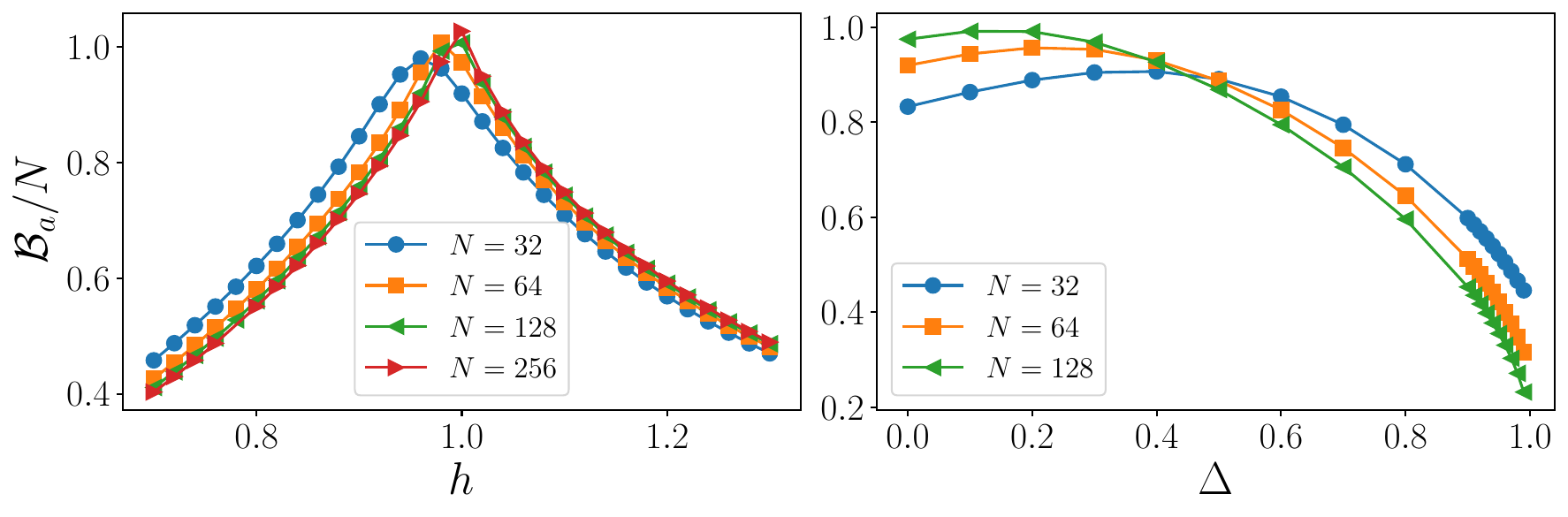}
        \put (0,35) {{\textbf{(a)}}}
        \put (50,35) {{\textbf{(b)}}}
        \end{overpic}
        \caption{The additive Bell magic density $\mathcal{B}_a/N$ for the ground state of (a) the quantum Ising chain as a function of the transverse field $h$ and (b) the XXZ chain as a function of the anisotropy $\Delta$.  }.
        \label{fig:bell_magic}
    \end{figure}

We perform benchmarking simulations of the additive Bell magic in the ground states of the Ising and XXZ chains, shown in Fig. \ref{fig:bell_magic} (a) and (b), respectively. We find that the additive Bell magic exhibits similar behavior to that of the SRE \cite{haug2023quantifying,haug2023stabilizer}. Moreover, we investigated the growth of the Bell magic under random Clifford circuits doped with a single $T$ gate per time step. Here the circuit is a brickwork of two–site Clifford gates chosen randomly from the set $\{ I, CNOT^{L},CNOT^{R} \}$. The initial state is polarized in the $y$ direction, and the $T$ gates are applied to a randomly chosen site at each time step. The results are shown in Fig. \ref{fig:additional} (a). We observe that, at short times, the additive Bell magic grows linearly with $t$.

Furthermore, we computed the Bell magic in a state prepared by a quantum circuit recently realized in Ref. \cite{Bluvstein2023}, for $N=9$. The relative circuit is shown in Fig. \ref{fig:circuit_L9}. Also in this case, we verify that the additive Bell magic increases as a function of the number of $CCZ$ gates applied. 

\subsubsection{Stabilizer nullity}
We present an additional result of random Clifford circuits for $N_T=1$ with circuit depth $D=N/4$ in Fig. \ref{fig:additional} (b). We again observe rapid convergence of $-\log_2 \norm{| P_k \rangle}$ to its expected value in all cases.
In addition to the case of random Clifford circuits, we calculated the stabilizer nullity in the context of ground states for the Ising and XXZ chains. For the Ising chain, the nullity is $\nu=N-1$ with stabilizer group $\{I_N, \prod_j \sigma^z_j\}$. For the XXZ chain, the nullity is $\nu=N-2$ with stabilizer group $\{I_N, \prod_j \sigma^x_j, \prod_j \sigma^y_j, \prod_j \sigma^z_j\}$. The results of our algorithm are shown in Fig. \ref{fig:additional} (c) for $N=128$, demonstrating that $-\log_2 \norm{| P_k \rangle}$ again reaches its expected value in both cases.

\subsubsection{Perfect sampling}
\begin{figure}
        \centering
        \begin{overpic}[width=0.35\linewidth]{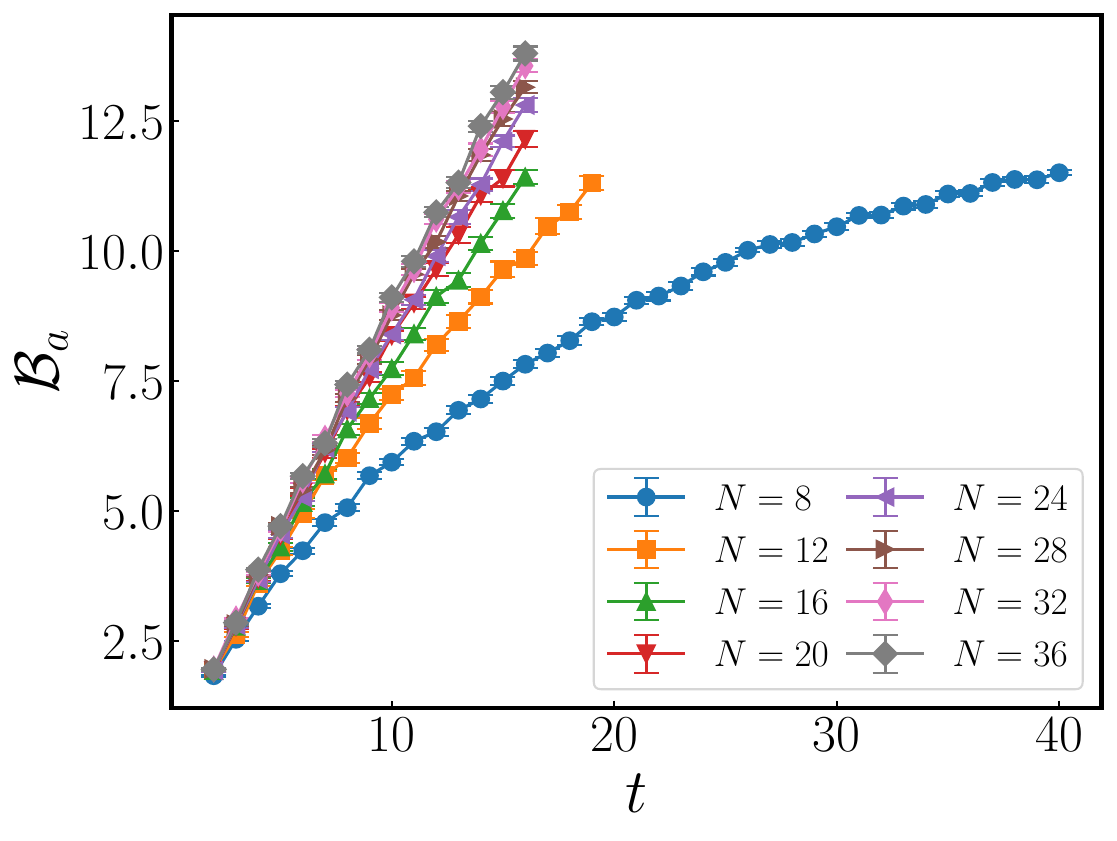}
        \put (-1,80) {{\textbf{(a)}}}
        \end{overpic}
        \begin{overpic}[width=0.35\linewidth]{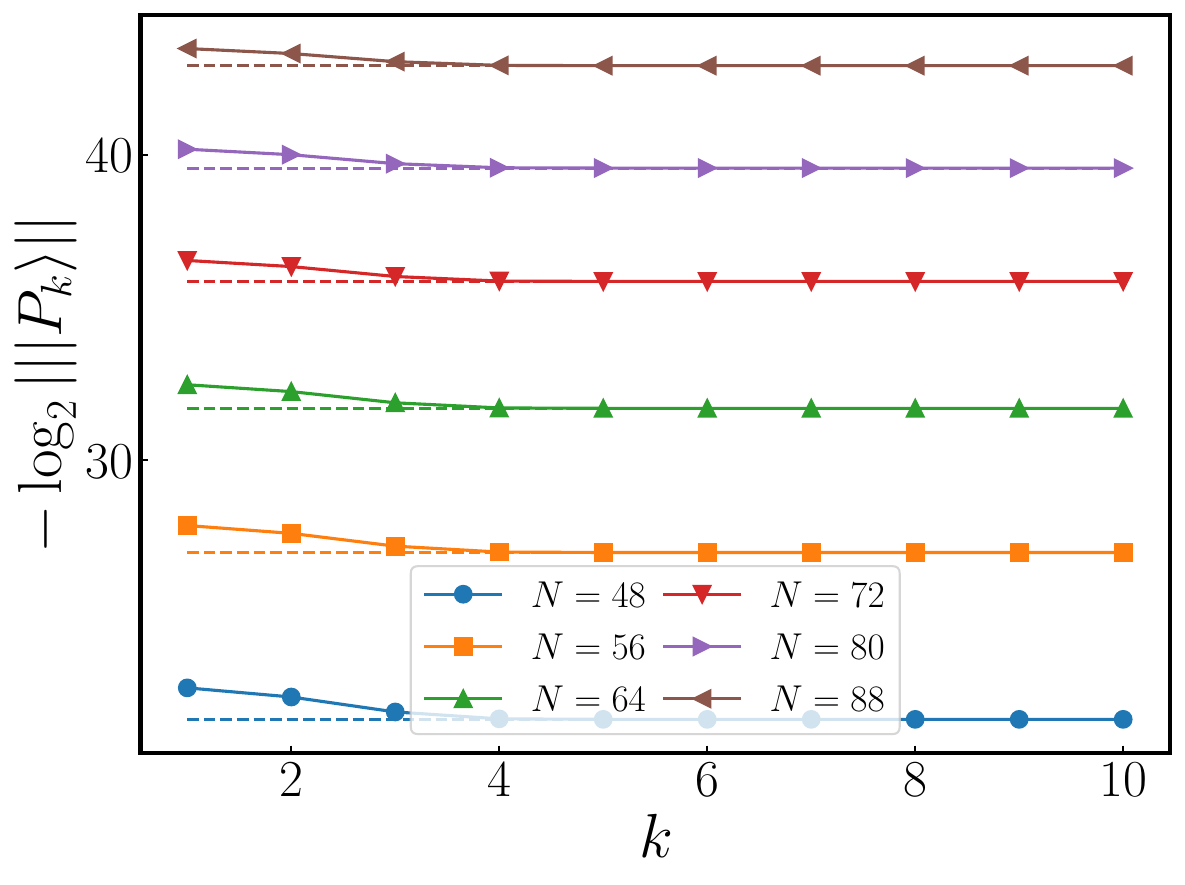}
        \put (-1,80) {{\textbf{(b)}}}
        \end{overpic}
        \begin{overpic}[width=0.35\linewidth]{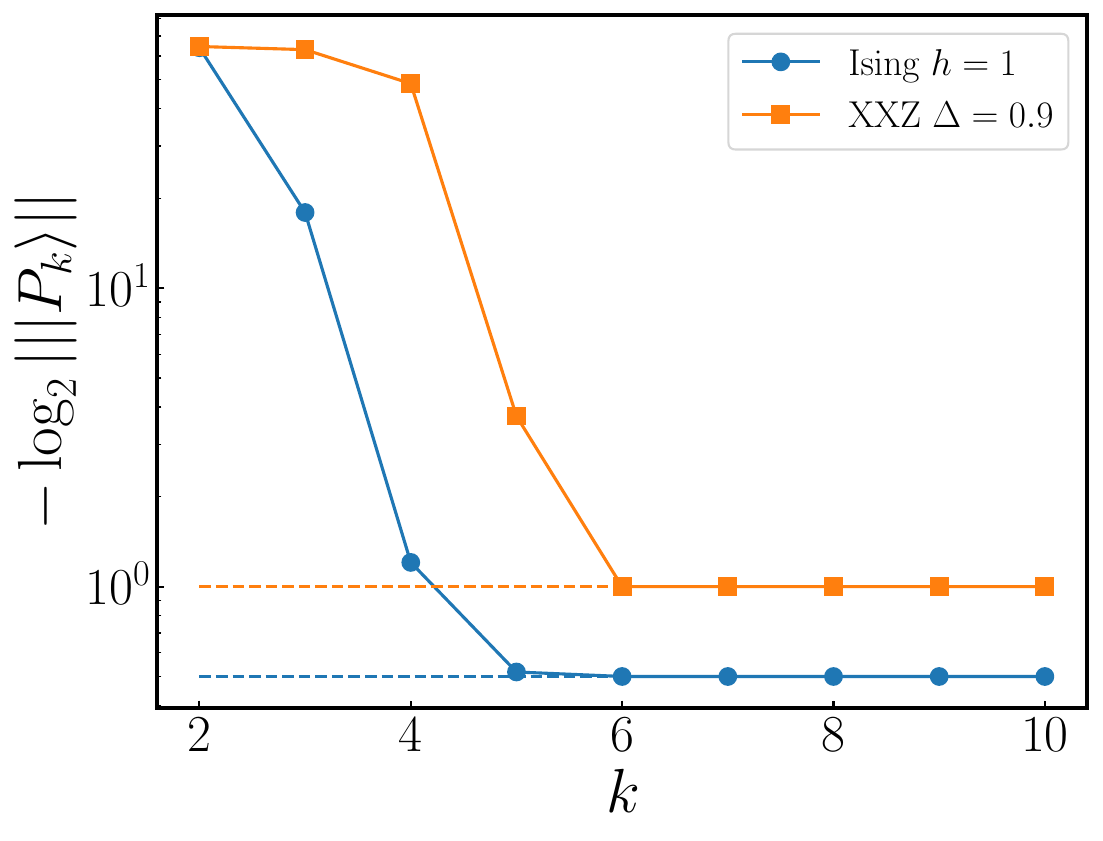}
        \put (-1,80) {{\textbf{(c)}}}
        \end{overpic}
        \begin{overpic}[width=0.35\linewidth]{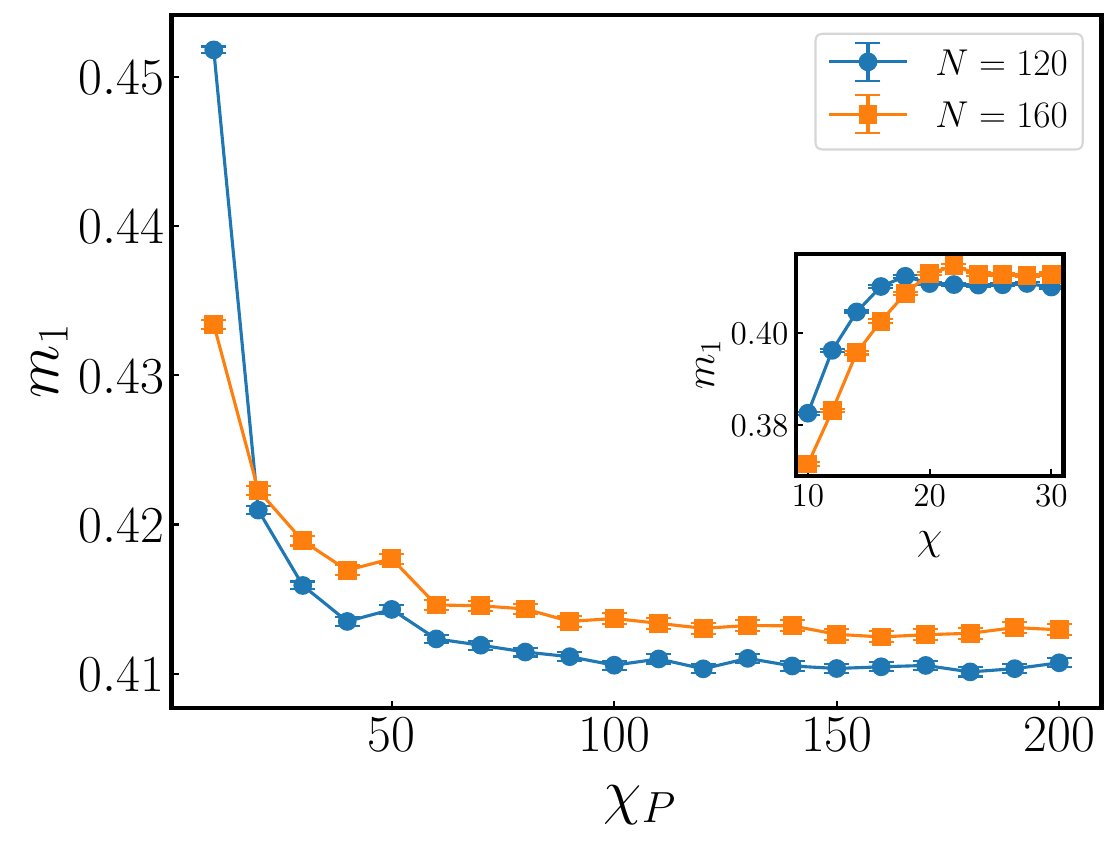}
        \put (-1,80) {{\textbf{(d)}}}
        \end{overpic}
        \caption{ (a) The additive Bell magic $\mathcal{B}_a$ in random Clifford circuits doped with a single $T$ gate per time step, averaged over 200 to 500 realizations. (b,c) $-\log_2 \norm{| P_k \rangle}$ at iteration $k$ (b) in the outputs of random Clifford circuits with $N_T$\,$=$\,$1$ number of $T$ gates and circuit depth $D$\,$=$\,$N/4$ as well as (c) in the ground state of the quantum Ising chain at the critical point $h=1$ and the XXZ chain at $\Delta=0.9$ with $N=128$. The dashed line denotes the analytically known $(N-\nu)/2$ for each system with the same color. (d) SRE density $m_1=M_1/N$ calculated with perfect sampling on $| P(\psi) \rangle$ as a function of $\chi_P$. The ground state is obtained with $\chi=60$. Inset: $m_1$ calculated by the perfect Pauli sampling as a function of $\chi$. Both results are for the ground state of the XXZ chain with anisotropy $\Delta=0.9$ and system size $N\in \{120,160\}$. The number of samples is $N_S=10^5$.}.
        \label{fig:additional}
    \end{figure}

Here, we show that our approach can also be applied to improve methods based on tensor network sampling, which recently have been proposed to estimate the SRE  ~\cite{Lami2023,haug2023stabilizer,tarabunga2023manybody}. In particular, the Pauli strings can be sampled directly according to the probability distribution $\Xi(\boldsymbol{\alpha})=|\langle \psi | P_{\boldsymbol{\alpha}} | \psi \rangle|^{2}/2^N$ via perfect Pauli sampling algorithm introduced in~\cite{Lami2023,haug2023stabilizer}. With the MPS representation in Pauli basis in Eq. \eqref{eq:pauli_mps}, this is equivalent to the perfect MPS sampling proposed in Ref.~\cite{Ferris2012} (see also Refs.~\cite{Stoudenmire2010,FrasPrez2023}). The cost scales as $O\left(\left(\chi^2\right)^2\right) = O\left(\chi^4\right)$ with respect to the bond dimension $\chi$. At first glance, this appears to be worse than the cost of perfect Pauli sampling in the MPS form of $|\psi \rangle$, which costs $O\left(\chi^3\right)$. However, similarly as in the replica method, we can truncate the bond dimension of the Pauli-MPS to a value $\chi_P$ considerably smaller than $\chi^2$, such that the perfect MPS sampling on Eq. \eqref{eq:pauli_mps} becomes superior compared to perfect Pauli sampling. This is particularly beneficial in computing the SRE of an MPS approximation of a state (e.g., ground states), where one can first obtain a well-converged MPS (e.g., by DMRG with large bond dimension), and truncate the bond dimension of its Pauli vector to a manageable value $\chi_P$. One can then study the accuracy of the computed SRE by varying $\chi_P$.  The comparison between perfect sampling in $| P(\psi) \rangle$ and perfect Pauli sampling in $| \psi \rangle$ is shown in Fig. \ref{fig:additional} (d). With our method, we find that $M_1$ can be converged with considerably less resources compared to the standard approach, even when accounting for the initial overhead of constructing the MPS in the Pauli basis.


\subsubsection{Comparison between replica Pauli-MPS and perfect Pauli sampling}

\begin{figure} 
    \centering
    \includegraphics[width=0.45\linewidth]{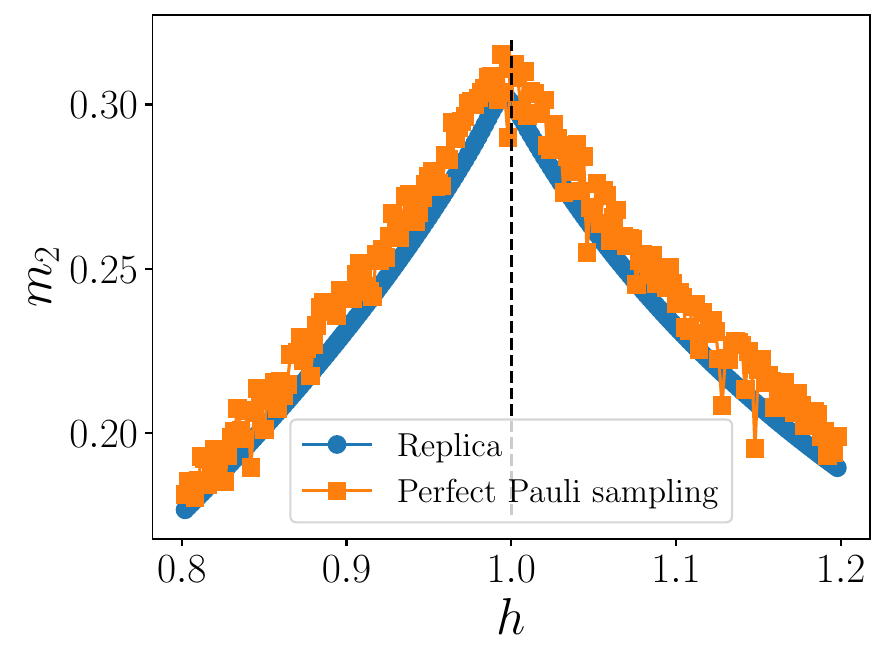}
    \caption{Comparison between the SRE density $m_2$ in the ground states of the transverse field Ising chain for $N=256$ obtained with our method and perfect Pauli sampling. The ground states are obtained using DMRG with bond dimension $\chi=40$ and $\chi_P=80$. The data for perfect Pauli sampling is obtained with $N_S=10^5$ samples.}
    \label{fig:comparisonL=256}
\end{figure}

We compare the SRE density $m_2$ obtained with replica Pauli-MPS method and perfect Pauli sampling for the ground states of the transverse field Ising chain with system size $N=256$ in Fig. \ref{fig:comparisonL=256} . We applied our replica method with $\chi=40$ and $\chi_P=80$. Compared to the more accurate data obtained with error threshold $\epsilon=10^{-9}$ presented in the main text, the discrepancy is at most on the order of $10^{-4}$. With perfect Pauli sampling, we estimated $m_2$ using $N_S=10^5$ samples. We see that the results of perfect Pauli sampling for $N=256$ display appreciable statistical errors. This is expected since the number of samples has to scale exponentially to maintain the precision for the estimation of $m_2$. Thus, this demonstrates the limitation of perfect Pauli sampling to scale up to larger systems. 

Note that it took at most 15 minutes to obtain each data point with our replica method, while perfect Pauli sampling took a few hours per data point (both codes were run on the same computing cluster). This highlights that the replica method surpasses perfect Pauli sampling not only in terms of accuracy and scalability, but also in terms of  computational efficiency. Moreover, the replica method is more powerful at computing derivatives or linear combinations of the SRE, which pose significant challenges for perfect Pauli sampling.

%% file: draft_prl.bbl
\begin{thebibliography}{98}
\bibitem [{footnote1()}]{footnote1}%
  \BibitemOpen
  \bibinfo {note} {Although the probability distribution $\Xi(\boldsymbol{\alpha})$ differs from the one employed in Ref. \cite{PRXQuantum.4.010301}, which utilizes the probability distribution of Pauli strings obtained from two-copy Bell measuremennts, both definitions of the Bell magic are mathematically equivalent.}\BibitemShut {Stop}%
\bibitem [{footnote2()}]{footnote2}%
  \BibitemOpen
  \bibinfo {note} {We note that, since $| P^{(n)}(\psi) \rangle$ is real, the computation of the norm does not require complex conjugation.}\BibitemShut {Stop}%
\bibitem [{footnote9()}]{footnote9}%
  \BibitemOpen
  \bibinfo {note} {The MPS $|Q(\psi) \rangle$, which has physical dimension 4 and exact bond dimension $\chi^8$, stores the probability distribution that can be obtained by Bell difference sampling \cite{Gross2021}.}\BibitemShut {Stop}%
\bibitem [{footnote5()}]{footnote5}%
  \BibitemOpen
  \bibinfo {note} {Our numerical data consistently show that the bond dimension is non-increasing after each iterations, although we were not able to provide a proof. This also implies that $\chi_{\mathrm{max}}$ never exceeds $\chi^2$.}\BibitemShut {Stop}%
\bibitem [{footnote6()}]{footnote6}%
  \BibitemOpen
  \bibinfo {note} {The initial iterations, where the bond dimensions are the largest, require the most computational resources. The bond dimensions grow smaller with each iteration, resulting in significantly faster computations in the subsequent iterations.}\BibitemShut {Stop}%
\bibitem [{footnote7()}]{footnote7}%
  \BibitemOpen
  \bibinfo {note} {One can see $|G(\psi) \rangle$ as the Pauli vector of $\rho^{(\infty)}$, whose Pauli expectation values are 1 if $\langle \psi | P_{\boldsymbol{\alpha}} |\psi \rangle$\,$=$\,$\pm 1$, and 0 otherwise. $\rho^{(\infty)}$ is thus a (normalized) projector onto the stabilizer group of $| \psi \rangle$.}\BibitemShut {Stop}%
\end{thebibliography}
